\pdfoutput=1
\documentclass[journal,12pt,onecolumn,draftclsnofoot,]{IEEEtran}

\usepackage[T1]{fontenc}

\usepackage{cite}
\usepackage{graphicx}
\usepackage[fleqn]{amsmath}
\usepackage[cmintegrals]{newtxmath}

\usepackage{comment}
\usepackage{bm}
\usepackage{amsfonts}

\usepackage{multirow}
\usepackage{booktabs}
\ifCLASSOPTIONcompsoc
 \usepackage[caption=false,font=normalsize,labelfont=sf,textfont=sf]{subfig}
\else
 \usepackage[caption=false,font=footnotesize]{subfig}
\fi
\usepackage{url}

\begin{document}
\title{Proactive Received Power Prediction Using Machine Learning and Depth Images for mmWave Networks}

\author{Takayuki~Nishio,~\IEEEmembership{Member,~IEEE,}
        Hironao~Okamoto,~\IEEEmembership{}
        Kota~Nakashima,~\IEEEmembership{Student Member,~IEEE,}
        Yusuke~Koda,~\IEEEmembership{Student Member,~IEEE,}
        Koji~Yamamoto,~\IEEEmembership{Member,~IEEE,}
        Masahiro~Morikura,~\IEEEmembership{Member,~IEEE,}
        Yusuke~Asai,~\IEEEmembership{Member,~IEEE,}
        and~Ryo~Miyatake~\IEEEmembership{Member,~IEEE,}%
\thanks{T.\ Nishio, H.\ Okamoto, Y.\ Koda, K.\ Nakashima, K.\ Yamamoto, and M.\ Morikura are with the Graduate School of Informatics, Kyoto University, Kyoto, 606-8501 Japan (e-mail: nishio@i.kyoto-u.ac.jp).}
\thanks{Y.\ Asai and R.\ Miyatake are with NTT Network Innovation Laboratories, NTT Corporation, Yokosuka, Kanagawa, 239-0847 Japan.}
\thanks{Parts of this work were presented at the 85th IEEE Vehicular Technology Conference (VTC Spring) and the IEEE Consumer Communications and Networking Conference (CCNC).}
\thanks{This work was supported in part by JSPS KAKENHI Grant Number 17H03266.}
}


\maketitle

\begin{abstract}
    This study demonstrates the feasibility of the proactive received power prediction by leveraging spatiotemporal visual sensing information toward the reliable millimeter-wave (mmWave) networks.
    Since the received power on a mmWave link can attenuate aperiodically due to a human blockage, the long-term series of the future received power cannot be predicted by analyzing the received signals before the blockage occurs.
    We propose a novel mechanism that predicts a time series of the received power from the next moment to even several hundred milliseconds ahead. 
    The key idea is to leverage the camera imagery and machine learning (ML). 
    The time-sequential images can involve the spatial geometry and the mobility of obstacles representing the mmWave signal propagation. 
    ML is used to build the prediction model from the dataset of sequential images labeled with the received power in several hundred milliseconds ahead of when each image is obtained.
    The simulation and experimental evaluations using IEEE 802.11ad devices and a depth camera show that the proposed mechanism employing convolutional LSTM predicted a time series of the received power in up to 500\,ms ahead at an inference time of less than 3\,ms with a root-mean-square error of 3.5\,dB.
\end{abstract}


\IEEEpeerreviewmaketitle

\section{Introduction}\label{sec:intro}

\IEEEPARstart{M}{illimeter-wave} (mmWave) communication technology has attracted considerable attention for the fifth-generation (5G) and the beyond mobile networks \cite{Dehos2014,Sakaguchi2015}.
The 5G and beyond networks will support high data rate and large volume traffic, very low latency, and high reliability for providing various applications such as cloud services, virtual reality (VR), and augmented reality (AR) \cite{Osseiran14, IMT15}. 
The mmWave bands can meet the demand, i.e., the mmWave band offers a large bandwidth on the order of several GHz, and enables multi-Gbit/s transmission rates.
For example, the IEEE 802.11ad standard using a 2.16\,GHz bandwidth in 60 GHz achieves 4.62\,Gbit/s with a single carrier transmission. 
However, the high pathloss and the sensitivity to blockage render it challenging to provide reliable wireless links. In particular, the received power suddenly decreases by 20\,dB or more when a line-of-sight (LOS) path is blocked by obstacles such as pedestrians and vehicles \cite{Collonge2004, yamamoto2008}.
Such a sudden and damaging attenuation can cause burst packet drops and temporal degradation of the transmission rate and throughput. This is critical for the 5G and beyond 5G applications, especially VR/AR, which require high data rates and low latency. 

A number of communication control schemes such as access point (AP) handover \cite{Gao2014,react1} and fast session transfer (FST) \cite{perahia2011gigabit} have been studied to solve the blockage problem and provide the reliable mmWave communications. These schemes reactively conduct the handover and FST based on the measured link quality, i.e., the operation is performed after the received power attenuates considerably. However, the blockage decreases the received power sharply; in particular, the mean time to decay by 10\,dB by the blockage is 39\,ms \cite{11ad_channel_model}. Therefore, considerable packet loss and throughput degradation will occur until the control operations are completed, as well as while during the operations.
The key to enabling the highly reliable communications using the mmWave band is a predictive network control that conducts operations before the link quality decreases. A novel predictive handover and beam switching are proposed in \cite{nishio15, oguma15, simic16}. In \cite{nishio15} and \cite{simic16}, a camera and mmWave radar are employed to detect a human approaching the LOS path, respectively. When the human reaches the vicinity of the LOS path, a control system predicts the human blockage and conducts flow control or beam switching to avoid the use of the blocked communication path. Oguma et al. proposed a predictive handover system based on human blockage prediction using a camera, where the mobility of pedestrians is predicted by leveraging human recognition techniques, and the timing when the blockage occurs is predicted explicitly \cite{oguma15}. The experimental evaluation confirmed that the system significantly reduces the amount of time when the throughput is degraded by human blockages compared with a reactive handover scheme.

As confirmed by the previous works, the predictive control based on the deterministic prediction enables the human blockage to be avoided proactively, and realizes reliable and high-throughput mmWave communications. 
However, these methods cannot predict the amount of signal attenuation by a blockage, though these methods can predict the time when the blockage may occur.
The attenuation varies depending on the environment such as the position blocked by the obstacle, and the shape and material of the obstacle. When the attenuation is not large, the controller may not be required to conduct a handover or beam switching. However, it is difficult to predict the attenuation because it depends on the situations, e.g., positions, shapes, and materials of obstacles and the existence of a strong reflected path, all of which dynamically changes the situation.
The latency of a handover procedure is several tens or hundreds of milliseconds in LTE \cite{ulvan13}; thus, the proactive handover is required to predict the received power at least several tens or hundreds of milliseconds before the blockage.

Moreover, for interactive VR/AR applications such as gaming, eSports, and VR education which require low latency, high capacity, and high reliability for communication among distant persons, the end-side proactive control methods such as video streaming control \cite{satoda12} and proactive content caching \cite{kanai16} must be important to avoid the degradation of the quality of experience induced by the blockage in mmWave communications. To predict the future received power by more than several hundred milliseconds before the blockage is required for enabling the end-side proactive control owing to the long round trip time.


The empirical and stochastic analysis of a mmWave channel with the blockage has been investigated to provide stochastic prediction models \cite{Haneda2015, Collonge2004, di2015stochastic, gapeyenko2016}. These prediction models enable us to determine the occurrence frequency of the blockage, and the impact degree of the blockage to the mmWave communications; however, it is impossible to determine when the blockage occurred and the amount of signal attenuated by the blockage. 
Some works studied deterministic prediction of a future link quality using time-series prediction models \cite{LQ_prediction,LQ_prediction2, yao08}. These works analyze time series of link quality such as the SNR and packet reception rate (PRR), and capture trends such as a periodic degradation of SNR and PRR induced by a periodic mobility. If some trends can be captured from the received power or wireless signals before a blockage occurs in the mmWave communications, the signal attenuation induced by the blockage could be predicted. However, the human blockage is typically aperiodic, i.e., a sign of the blockage is non-existent until the received power starts to decay. Therefore, it is difficult to predict a long-term time series of the future received power where the attenuation by the blockage exists from wireless signals.


This paper demonstrates the feasibility of the accurate prediction of the future mmWave received power after several hundred milliseconds, which is difficult for the above mentioned approaches. The key idea is to leverage camera images and machine learning (ML). 
The depth camera images can involve the geometry and the dynamics of the communication environment, i.e., the geographical relationships and the mobility of obstacles, which represent wireless signal propagation, especially mmWave signals. 
We regard the received power prediction problem as a regression problem in supervised learning, and ML algorithms are applied to learn the mapping from the depth images to the received power values.
This paper expands our previous work \cite{okamoto_vtc, okamoto_ccnc} to quantitatively predict the time series of received power including the future (i.e., several hundred milliseconds ahead) on a mmWave link. 
In \cite{okamoto_vtc, okamoto_ccnc}, we proposed current throughput or received power estimation schemes from depth camera images. The schemes enable the estimation of the throughput or received power at the time when the image is obtained on a mmWave link even when an access point (AP) and a station (STA) are not communicating at the time. However the previous schemes cannot predict the future throughput and received power.
This paper extends the previous works to predict the future received power by employing a novel data preprocessing technique. 
The contributions of this paper are summarized as follows:
\begin{itemize}
    \item The feasibility of the accurate prediction by leveraging the time-sequential depth images is demonstrated by experimental evaluations using the dataset obtained from simulations and that from mmWave experiments. In the mmWave experiments, the received power values are measured from the mmWave signals by using commercially available IEEE 802.11ad devices, and the depth images are captured by an \mbox{RGB-D} camera Kinect.
    To our best knowledge, this is the first work to predict the mmWave received power even up to several hundred milliseconds ahead in the environment where the blockage occurs aperiodically.
    \item In order to enable ML algorithms to accurately predict the future received power, we newly design a data preprocessing technique that labels the time-sequential images with the future received power values in several hundred milliseconds ahead of the time when the latest image in the sequential image is captured.
    \item We design two examples of structures of neural networks which predicts the received power at an inference time of less than several milliseconds with the root-mean-squre (RMS) errors of less than several dB.
    Experimental results show that all of the ML algorithms can predict the received power from depth images. The results also show that the designed NN employing convolutional long short-term memory (ConvLSTM), which can consider spatiotemporal features, achieves the highest accuracy among the algorithms. 
\end{itemize}

This paper is organized as follows.
The related works are introduced in Section \ref{sec:related}. Section \ref{sec:ml_based_lq_prediction} describes the proposed ML-based received power prediction mechanism.
Section \ref{sec:ml_algorithm_and_model} introduces the ML algorithms used for the received power prediction.
Sections \ref{sec:simulation} and \ref{sec:experiments} shows the experimental evaluations using a dataset obtained from the simulation and that from mmWave experiments, respectively.
Conclusions are presented in Section \ref{sec:conclusion}.

\section{Related Works}\label{sec:related}
As mentioned in Section \ref{sec:intro}, the empirical and stochastic analyses of the mmWave channel with the blockage have been well investigated \cite{Haneda2015, Collonge2004, di2015stochastic, gapeyenko2016}. These works focus on providing stochastic prediction models. Maltsev et al. investigated a probabilistic model of human blockage and experimentally confirmed that a mean attenuation induced by the human blockage follows the Gaussian distributution, whereas the duration of the blockage follows the Weibull distribution \cite{11ad_channel_model}. However, these models cannot exactly predict when the next blockage will occur, and the amount of signal attenuation; nevertheless, these models enable us to estimate the occurrence frequency of the blockage, and the expected amount of attenuation to the received power.

A time-series analysis is a straightforward approach to predict time-series data \cite{zhang03}. Generally, the time-series analysis can predict periodic and long-term trends, but it cannot predict aperiodic variation. In \cite{LQ_prediction, LQ_prediction2, yao08}, time-series of link qualities such as SNR, PRR, and capacity was predicted in conventional microwave communications. The periodic and long-term patterns of the link quality are captured to predict the future link quality. ML-based prediction methods using time-series data have also been studied in \cite{RNN_SNR, long08}. As like time-series analysis, these methods learn periodic and long-term trends to predict the received power by using recurrent neural networks (RNNs).
However, as mentioned in Section \ref{sec:intro}, a human blockage occurs aperiodically in mmWave communications and there are no signs of the blockage several hundred milliseconds before the blockage occurs. Therefore, although it is possible to reactively estimate the signal attenuation induced by the human blockage, it is difficult to predict the human blockage before it occurs  

Another approach to predict the received power is a ray-tracing simulation \cite{ray_tracing_11ad,ray_tracing_mmwave_calc,double_knife_edge}.
If a perfect geometry of a future communication environment, i.e., future positions, shapes, and materials of all the things in the environment (e.g., pedestrians, furniture, and walls), can be obtained, the ray-tracing simulation can calculate all the possible signal propagation and predict the received power. However, it is difficult to obtain such information accurately. Moreover, for accurate ray-tracing simulations, a large number of rays including reflected and diffracted rays should be calculated, which requires significant computational time to predict. According to \cite{ray_tracing_mmwave_calc}, more than 200\,ms is required to simulate one frame of mmWave propagation considering only two reflections with a 2\,GHz CPU. The computation time is larger than a general beacon interval of the IEEE 802.11ad mmWave WLANs, where TXOP scheduling and beam alignment are conducted \cite{11ad}.

A camera-assisted proactive blockage prediction and network control system for mmWave networks have been proposed in \cite{nishio15,oguma15,oguma2016proactive}.
This system captures the mobility of obstacles using depth cameras and conducts network operations such as flow control and handover proactively before the blockage occurs.
In the system, pedestrians are detected from depth images and localized by computer vision techniques such as those reported in \cite{conv3d, biswas2012depth}. The pedestrian velocity is calculated from the time-series of its location, and the time when the pedestrian enters an area where an LOS path exists is predicted.
Simic et al. proposed a radar-assisted MAC protocol for mmWave networks that detects obstacles approaching the LOS path using radars \cite{simic16}. When an AP detects the obstacles existing in a predetermined area, the AP conducts beam switching to avoid transmitting frames on the path that would likely be blocked by the obstacles. 
These sensor-assisted approaches enable us to proactively predict the link quality without transmitting any data and sensing frames between the AP and the STAs, and the delay for the prediction is smaller than 100\,ms at the most. However, as mentioned in Section \ref{sec:intro}, these methods cannot exactly predict the received power values that are helpful for network controllers to manage mmWave links precisely. Moreover, the camera-based prediction fails to predict when the human detection fails.


\section{Machine-Learning-Based Received Power Prediction}
\label{sec:ml_based_lq_prediction}
\subsection{System model}

\begin{figure}[!t]
    \centering
    \includegraphics[width=0.46\textwidth]{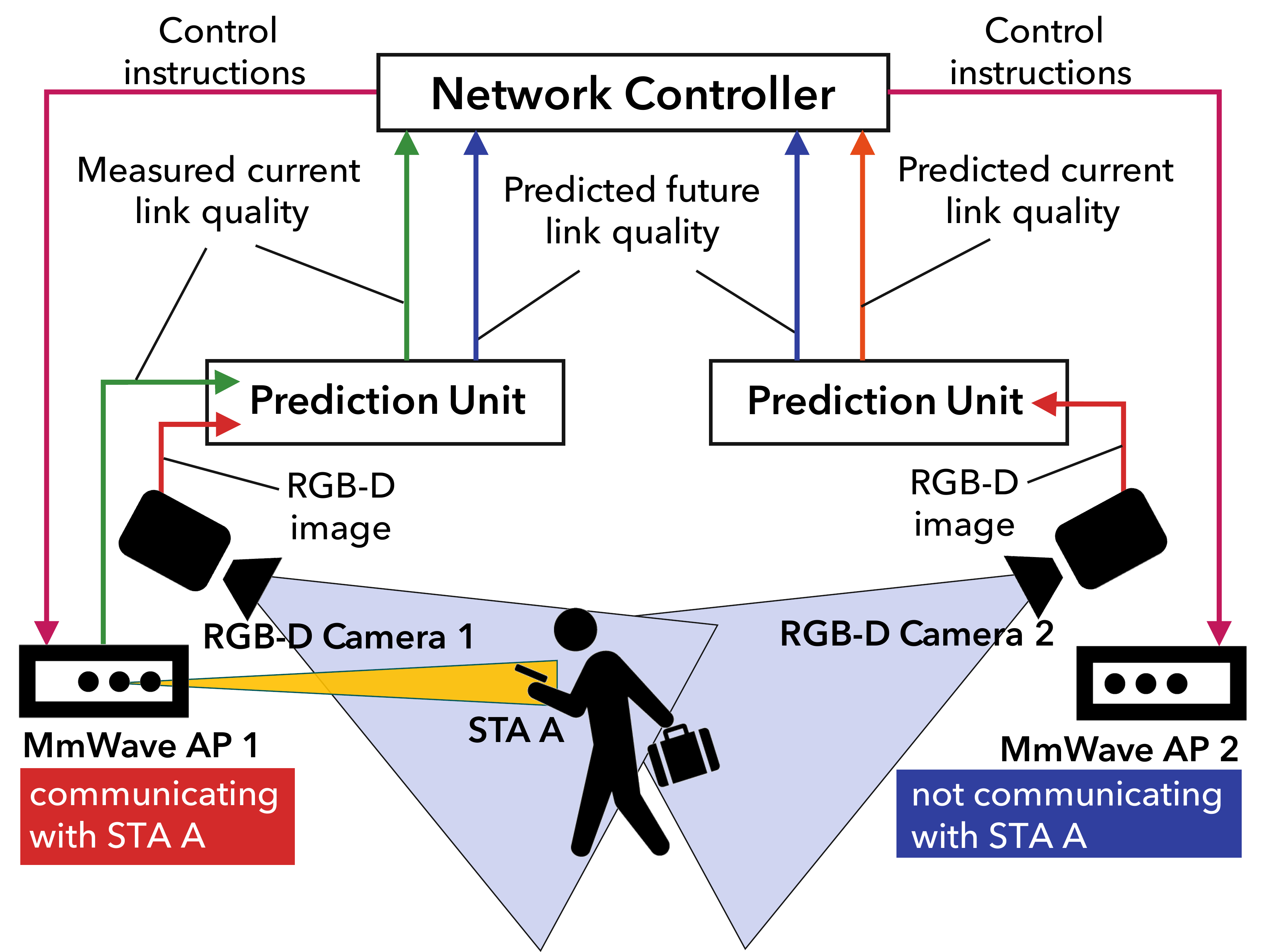}
    \caption{System model. The prediction units predict the link quality from the depth images. The network controller conducts network operations based on both the predicted and measured link quality.}
    \label{fig:scheme}
  \end{figure}

The system model of the proposed mechanism is consistent with the previous work \cite{nishio15, oguma15, oguma2016proactive}. 
Fig. \ref{fig:scheme} presents the system model. The system consists of mmWave APs, RGB-D cameras, prediction units, and a network controller.
Each prediction unit is connected with the mmWave AP, RGB-D cameras, and the network controller via wired local area networks.
RGB-D cameras, typified by the Microsoft Kinect \cite{kinect}, capture not only color images, but also depth images. 
Note that depth information can be obtained from RGB cameras using image processing techniques such as stereo vision techniques \cite{lazaros2008review}.
RGB cameras are widely available as surveillance cameras in crowded places such as airports and commercial facilities, which are primary targets for mmWave communications deployment. Therefore, the cameras can be shared by other applications such as a surveillance, a user localization, and a navigation in order to reduce an initial and running cost to deploy cameras.
In this paper, only depth images are used for prediction, since these camera-to-object distance data in depth images reflect the mobility of obstacles and users, which significantly affects mmWave link quality.
The cameras send the RGB-D images to the connected prediction units.
Note that the prediction units may be connected to multiple cameras and can share cameras with other prediction units.
Unlike \cite{oguma2016proactive}, the prediction units not only predict whether the LOS will be blocked but also predict the received power values from the images.

In this paper, we assume a simple case where an STA does not move and the LOS path between the STA and the AP exists in the field of view (FOV) of the RGB-D camera.
A case where multiple STAs move around will be included in our future work.

\subsection{Learning and Prediction Procedure}
The operations of the proposed ML-based prediction are roughly divided into learning and prediction.
The two different operations depend on the availability of the measured received power.
When the measured received power is available, i.e., the STA transmits signals on the mmWave link and the AP received them, the learning is operated.
The prediction unit obtains a time series of the measured received power from the AP and a time series of depth images from cameras. The unit trains and updates the ML model by using the training dataset of the measured received power values and the depth images. A detailed data processing and the ML algorithm used for the prediction are described in the following sections.
When the measured received power is not available, the prediction unit cannot update the ML model.

After the ML model is trained, the prediction can be operated whenever a time series of depth images is obtained.
The prediction unit predicts the future received power using a currently obtained time series of depth images and the trained ML model. The detail is described also in the following section.
The prediction results are sent to the network controller.
The network controller conducts a network control based on the measured and/or predicted received power provided by the prediction units.

\subsection{Data Preprocessing for Received Power Prediction}\label{sec:preprocessing}
In the learning operation, the RGB-D camera captures depth images at a certain frame rate $\mathrm{F}$ (e.g., 30\,fps in Kinect), and the AP simultaneously captures the received power at the STA.
The prediction unit generates a training dataset from these received power and images.

\begin{figure}[!t]
  \centering
  \includegraphics[width=0.48\textwidth]{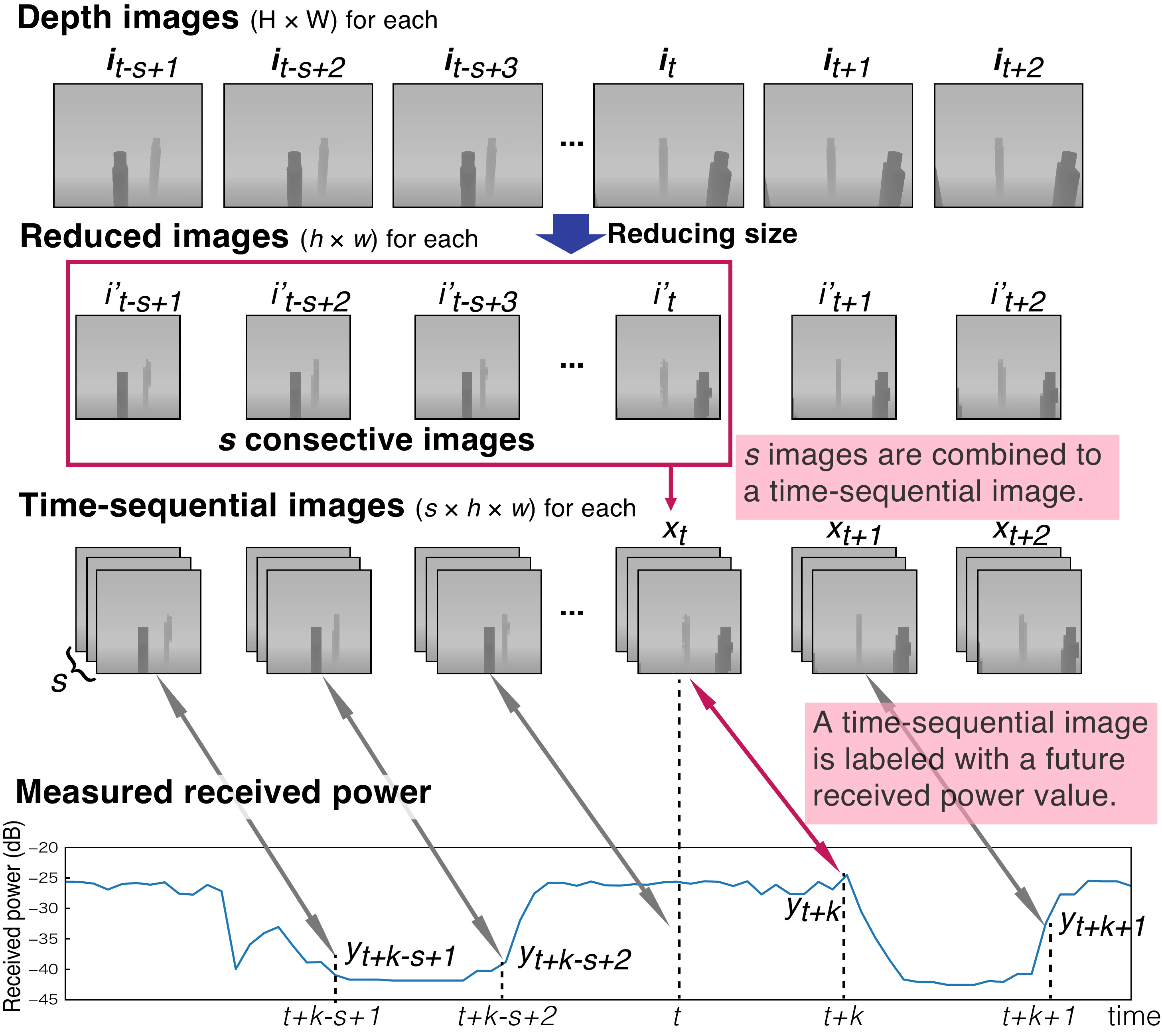}
  \caption{Dataset generation procedure. The depth images are reduced, combined, and associated with the future received power to generate a training dataset.}
  \label{fig:procedure}
\end{figure}

Fig.~\ref{fig:procedure} shows the dataset generation procedure for learning.
Let $t \in \mathbb{Z}$ denote a time index and $y_{t}$ be the measured received power obtained from an mmWave AP at $t$. A depth image is represented by a matrix with shape $\mathrm{H} \times \mathrm{W}$, where $\mathrm{H}$ and $\mathrm{W}$ are height and width of the depth image. Let $\bm{i}_{t}$ be the depth image obtained from the camera also at $t$. The received power and images are obtained at $\mathrm{F}$\,fps in the system, thus the system obtains $\mathrm{F}$ samples per seconds.
Firstly, in order to decrease calculation complexity, the size of each image is reduced to $h \times w$, where $h \leq \mathrm{H}$ and $w \leq \mathrm{W}$. Let $\bm{i}'_{t}$ be the reduced depth image.

Then, $s$ consecutive images are stacked into a third-order tensor with shape $s \times h \times w$. The tensor represents a time-sequential image. 
Let the time-sequential image is denoted as $\bm{x}_{t} = \bigl[\bm{i}'_{t-s+1}, \bm{i}'_{t-s+2}, \cdots, \bm{i}'_{t} \bigr]$. 
The time-sequential image has geometric information up to $(s-1)/\mathrm{F}$ sec ago.
Thus, the time-sequential image can contain temporal dynamics of the environment, i.e., mobility of pedestrians and other obstacles, in addition to the spatial features of the environment.
The idea of using time-sequential image as an input of the ML algorithms has been applied for various ML tasks where objects move around and it helps to solve the tasks \cite{convLSTM, mnih2015}.


Next, the time-sequential images are labeled with values of the received power to generate training dataset. In our previous works \cite{okamoto_vtc, okamoto_ccnc}, an image is associated with a received power measured simultaneously with the image the image.
However, in the proposed mechanism, all the time-sequential images are associated with future received power values. Specifically, to predict the received power $k$ frames ahead (equal to $k/\mathrm{F}$\,s ahead), $\bm{x}_{t}$ was associated with $y_{t+k}$. We call this labeling technique as temporal difference labeling. To label the images with the future received power enables ML algorithms to learn a mapping from the image to the future received power directly.

Let $T$ denote an index set of $t$ for captured samples. 
The constructed training dataset is $D = \{\,(\bm{x}_{t}, y_{t+k}) \mid t \in T\,\}$.
The prediction unit trains its ML model using the training dataset in order to update the model and increase prediction accuracy.

For the received power prediction, the prediction unit requires at least $s$ consecutive images including an image obtained at the current time. Therefore, the unit stores the $s$ latest depth images so that the unit can predict the future received power at any time. When the prediction unit receives a new depth image to predict the future received power, the unit drops the oldest image from its image buffer, reduces the size of the image to $h \times w$, and adds the reduced image to the buffer. Then the unit combines the $s$ latest depth image into a time-sequential image and inputs the time-sequential image to the trained model to get a prediction result of the received power value at $k/\mathrm{F}$\,s later.

The preprocessing procedure has hyperparameters, $k$ and $s$.
$k$ is determined by the requirements of network control operations. When $\mathrm{F}=30$ and the operation requires to predict the received power 100\,ms ahead, $k$ is set to 3. 
On the other hand, $s$ is the hyperparameter to determine the prediction accuracy and the computation cost. Large $s$ enables ML algorithm to capture long-term trends, which may require to predict the received power accurately, but it enlarges the size of the input which increases parameters of ML models and the computation cost. Therefore, $s$ should be set in consideration with the tradeoff between the accuracy and computation cost. However, it is difficult to determine the optimal value of $s$ since the optimal value depends on the allowable computation time and the accuracy that can be achieved by the ML algorithms, which is unpredictable generally. Therefore, in this paper, $s$ was determined empirically, and other hyperprameters of ML algorithms, which are explained in the next section, were also determined empirically. The hyperprameters were tuned manually so that the proposed mechanism predicts the received power with low errors in reasonable computation time for the validation data in the experimental evaluation. $s$ was specifically set to $16$, which corresponds to images obtained in 0.5\,s by a 30\,fps camera.

\section{Machine Learning Algorithms}
\label{sec:ml_algorithm_and_model}
This section introduces the ML algorithms employed in our received power prediction scheme.
Many ML algorithms have been developed and the state-of-the-art deep learning outperforms conventional ML algorithms in some learning tasks, especially in computer vision tasks.
ML algorithms, especially deep neural networks have many hyperparameters, e.g., the structures, the type and the number of layers, and the number of units. 
The hyperparameters change the performance of the ML algorithms, therefore they should be decided carefully.

In the received power prediction from images, it is important to capture spatiotemporal features involved in the time-sequential image since the signal attenuation induced by a blockage is a spatiotemporal event.
We used three ML algorithms which can capture spatial and/or temporal features involved in images as candidates of ML algorithms in the proposed mechanism.
Specifically, these algorithms are convolutional neural network (CNN), convolutional LSTM (ConvLSTM), and random forest (RF), which are often used for computer vision tasks \cite{cnn_survey}. 
Although there are many other possibilities of ML algorithms more suitable for the received power prediction,
the comprehensive investigation of the ML algorithms and hyperparameter tunings are out of scope, as the focus of this work is in designing the mechanism to predict the mmWave received power and demonstrating its feasibility.

\subsection{General introduction of ML algorithms}
\subsubsection{Convolutional Neural Network}
The CNN, which employs convolutional layers, is one of the most successful deep learning models. It has achieved competitive performances in many visual tasks \cite{cnn_survey}.
In 2D CNN, convolutions are applied on 2D images. These extract features implied in spatial information.
Furthermore, the 3D CNN extracts features from both spatial and temporal dimensions. Ji et al. demonstrated that the 3D CNN works well for a task of human action recognition \cite{conv3d}. 
In the received power prediction, the human action is important factor since it causes a blockage. Therefore, the 3D CNN is expected to work well for the received power prediction from images as it works for the human action recognition.

\subsubsection{Convolutional LSTM}
The LSTM network \cite{LSTM} is a widely used recurrent neural network (RNN) architecture.
RNN can learn the time-dependent relationships among inputs and outputs. Hence, it has been employed for sequential modeling and prediction problems in wireless communications and computer vision \cite{RNN_SNR, video_classification}.
The ConvLSTM network \cite{convLSTM} is an extension of the LSTM architecture, which has convolutional structures and enables us to handle not only spatial but also temporal sequence prediction problems. 
Shi et al. demonstrated that a ConvLSTM network learns longer-term spatiotemporal features and achieves a better prediction performance than a conventional LSTM. For detail description of the algorithm, please refer to reference \cite{convLSTM}. Therefore, the ConvLSTM is also expected to work well for the received power prediction from images.
For detail description of the algorithm, please refer to reference \cite{convLSTM}.

\subsubsection{Random Forest}
RF \cite{random_forest} is one of the most typical ensemble learning models.
RF consists of many simple decision trees using a bootstrap sample of the data and a randomly selected subset of input features at each split while growing each tree.
Every tree predicts its output from an input vector, and the model outputs the mean prediction of these outputs.
Thus, RF has the advantage of two ML algorithms; bagging \cite{bagging} and random variable selection. This leads to a stable and accurate model.
A well-known application of this model is 3D location prediction of individual body parts using an RGB-D image \cite{Shotton2011}. We expect that the RF captures spatiotemporal features and predicts the received power from depth images as like predicting 3D location of body parts from RGB-D images.

\subsection{Structures and Hyperprameters of Machine Learning Models}

\begin{figure}[!t]
  \centering
  \subfloat[Overview of NNs. \label{fig:nn_detail}]{
    \includegraphics[width=0.45\textwidth]{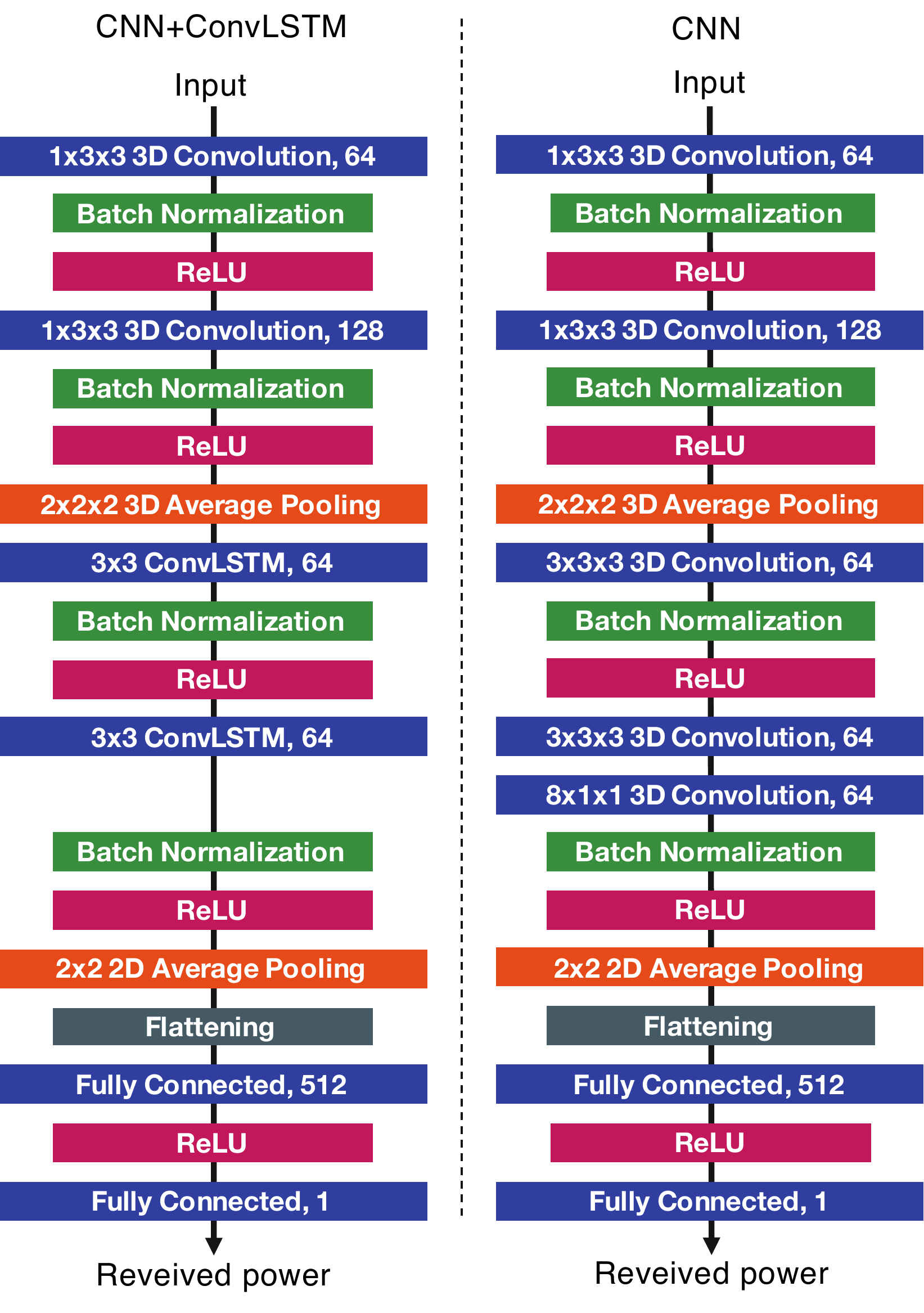}
  }\\
  \subfloat[Structure of ConvLSTM layers in CNN+ConvLSTM network. The first ConvLSTM layer returns spatiotemporal feature maps, while the second ConvLSTM layer returns spatial maps. \label{fig:convLSTM}]{
    \includegraphics[width=0.45\textwidth]{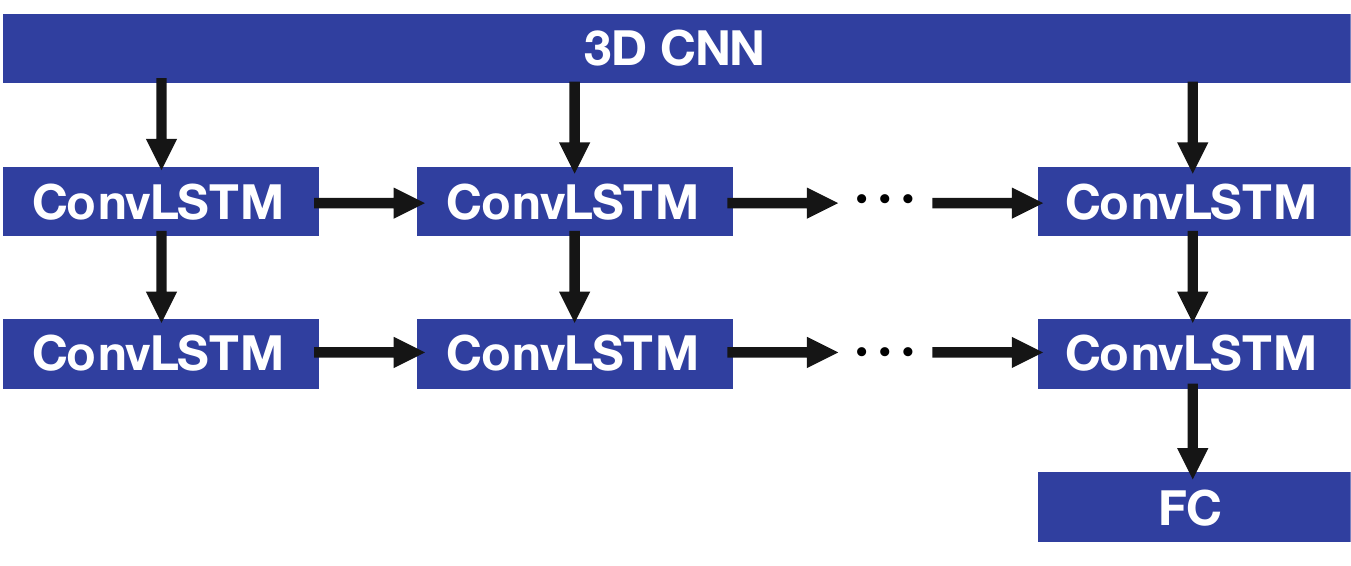}
  }
  \caption{Structures of NNs used in evaluations. \label{fig:nn}}
\end{figure}

We constructed two NNs called CNN+ConvLSTM and CNN, and employ RF as the candidates of the ML algorithms in the proposed mechanism.
Fig.~\ref{fig:nn} presents the structures of CNN+ConvLSTM and CNN. 
As mentioned in Section \ref{sec:preprocessing}, the structures and hyperprameters of NNs are tuned manually so that they predicts the received power from $s=16$ consecutive depth images with low errors for the validation data in the experimental evaluation.
These NNs take a three-dimensional input $\bm{x}_{t}$ and return a received power $y_{t+k}$.
They consist of several layers: 3D convolution (3D Conv) layers, ConvLSTM layers, Fully Connected (FC) layers, and average pooling layers.
Each layer gets its input from the previous layer and feeds its output to the next layer.
The average pooling layers reduce the size of feature maps by averaging the values that reduce calculation cost.
The FC layers are used to predict the received power from the feature map output from the previous layer.
The NNs employ batch normalization (BN) \cite{batch_normalization}, Rectified Linear Unit (ReLU) activation, and a flattening operation.
BN allows us to use much higher learning rates and be less careful about initialization of parameters of the NNs in order to accelerate training.
ReLU is the most widely used activation function. It is used to gain the non-linearity of the networks.
The flattening operation transfers a tensor to a vector to input it to the FC layer.

The first two 3D Conv layers adopt 64 and 128 convolution kernels of size $1 \times 3 \times 3$ (time $\times$ height $\times$ width).
The kernel size of time domain was set to 1 so that these 3D Conv layers are utilized to extract spatial features of each depth image in a time-sequential image.
The ConvLSTM layers adopt 64 convolution kernels of size $3 \times 3$  (height $\times$ width).
The first ConvLSTM layer returns 3D feature maps, while the second ConvLSTM layer returns 2D maps as shown in Fig.~\ref{fig:nn}\subref{fig:convLSTM}.
In the CNN, we employ three 3D Conv layers which have 64 kernels of size $3 \times 3 \times 3$, $3 \times 3 \times 3$, and $8 \times 1 \times 1$, in order. 
This is instead of the preceding ConvLSTM layers where the final output size of these three convolution layers is consistent with that of the two ConvLSTM layers.
While these two ConvLSTM layers or three 3D Conv layers are utilized to extract spatiotemporal features of each depth image, it is expected that the ConvLSTM layers can extract longer-term temporal features than the 3D Conv layers.

BN and a ReLU activation are applied after every ConvLSM and 3D Conv layer, except the fourth 3D Conv layer in the CNN.
The pool size of the average pooling layer is $2 \times 2 \times 2$.
This means each dimensional size of the feature map is reduced by half.
Then, the spatial feature maps are reduced to one-dimensional feature vectors in the flattening operation and the vectors are fed to the FC layers.
The unit number of the two FC layers are 512 and 1, in order, and the first FC layer is followed by ReLU activation.

RF can only take one-dimensional inputs, so a three-dimensional input $\bm{x}_{t}$ is reduced to one-dimensional feature vectors before the input is fed to the RF algorithm.
The other hyperparameters of the ML algorithms are shown in Table~\ref{tab:param}. 
We used Keras \cite{keras} with TensorFlow \cite{tensorflow} backend and scikit-learn \cite{sklearn} for the implementation of the abovementioned NNs and RF, respectively.
For an optimizer of NNs, we use Nadam \cite{nadam} with an initial learning rate of 0.001 and the other arguments set by default values in TensorFlow.
The learning rate is dropped to its $0.975$ every iteration.

\begin{table}[!t]
\centering
\caption{Hyperparameters of the ML algorithms.}
\label{tab:param}
\begin{tabular}{ccc}
\toprule
ML algorithm                           & Hyperparameter                 & Value              \\ \midrule
\multirow{3}{*}{\shortstack{CNN \\ \&\\ CNN+ConvLSTM}}             & batch size                & 64                 \\
                                        & epochs                    & 200                \\
                                        & loss function             & mean squared error \\\midrule
\multirow{3}{*}{RF}                     & number of trees           & 20                 \\
                                        & maximum depth of the tree & 20                 \\
                                        & splitting criterion       & mean squared error \\ \bottomrule
\end{tabular}
\end{table}

\section{Evaluation Using Simulation Data}
\label{sec:simulation}
The prediction accuracy of the received power prediction is evaluated by ML experiments using two datasets: dataset obtained from simulations and that from mmWave experiments.
This section introduces the evaluation using the simulation data.
We first performed ray-tracing simulations and 3D CG generations to obtain the training and test datasets for the received power prediction, and generated many datasets of situations where the ML algorithm and a camera position are varied.
Subsequently, the generated datasets are used for the ML experiments.
In the ML experiments, the dataset was split into training and test sets with a ratio of 80\% and 20\%, respectively.
A GeForce GTX 1080 Ti GPU was used to train each NN. The Intel Xeon CPU E5-1620 (3.50 GHz) was used to train the RF and to predict the received power.
When training the NNs, 25\% of the training set is used for a holdout validation to tune the structures and hyperparameters of the ML algorithms.

\subsection{Simulation Setup}

\begin{figure}[!t]
  \centering
  \includegraphics[width=0.24\textwidth]{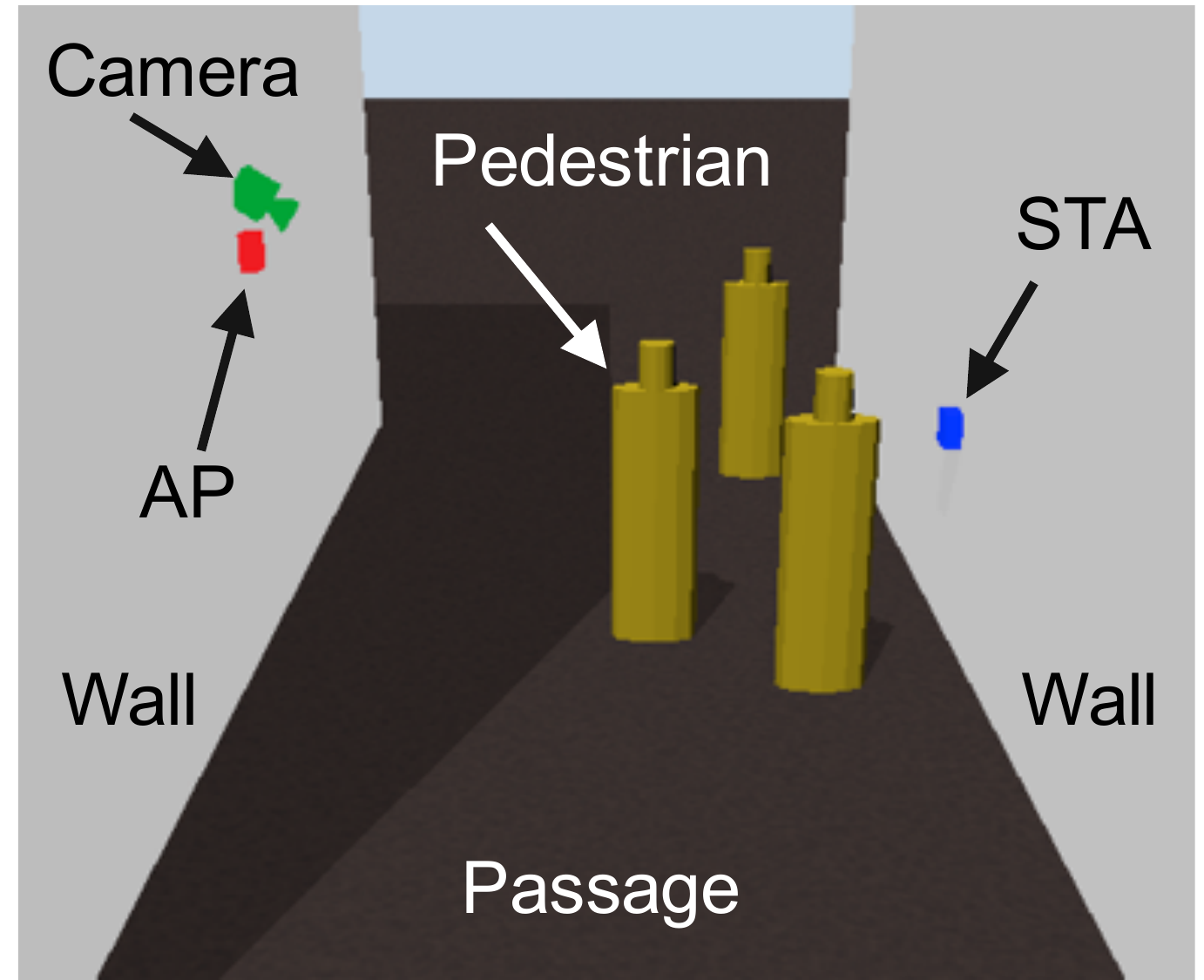}
  \caption{3D spatial model. The camera is placed at position $\mathrm{A_{low}}$ in Fig. \ref{fig:passage}.}
  \label{fig:camera_image_blender}
\end{figure}

\begin{figure}[!t]
  \centering
  \includegraphics[width=0.4\textwidth]{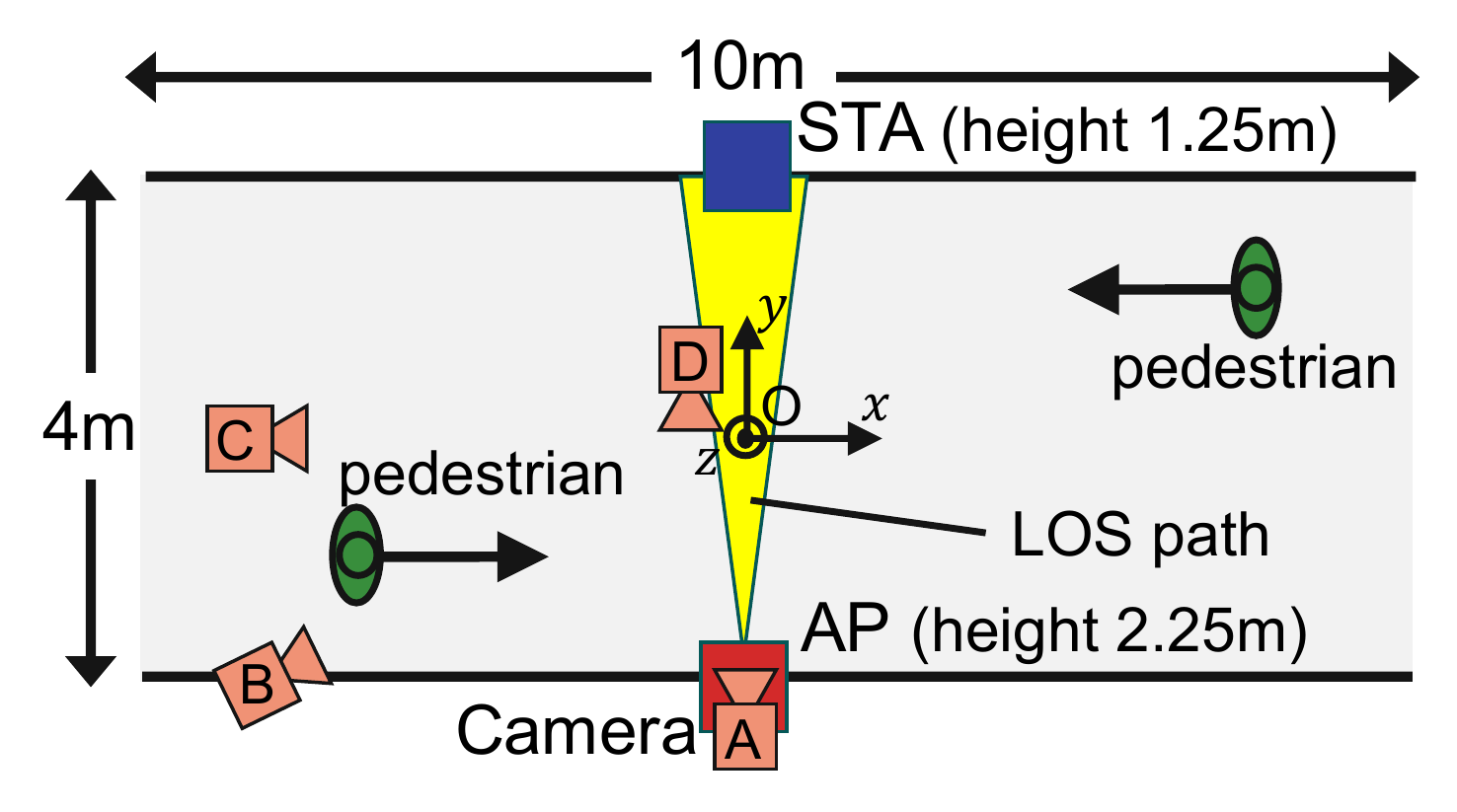}
  \caption{Passage model. The width of the passage is 4\,m and the length is 10\,m. Pedestrians walk on the right-hand side of the passage, from their perspective. A camera is placed at one of the eight positions: A--D in the X-Y plane and at a height of 2.25\,m (low) or 5\,m (high).}
  \label{fig:passage}
\end{figure}

\begin{figure}[!t]
  \centering
  \includegraphics[width=0.23\textwidth]{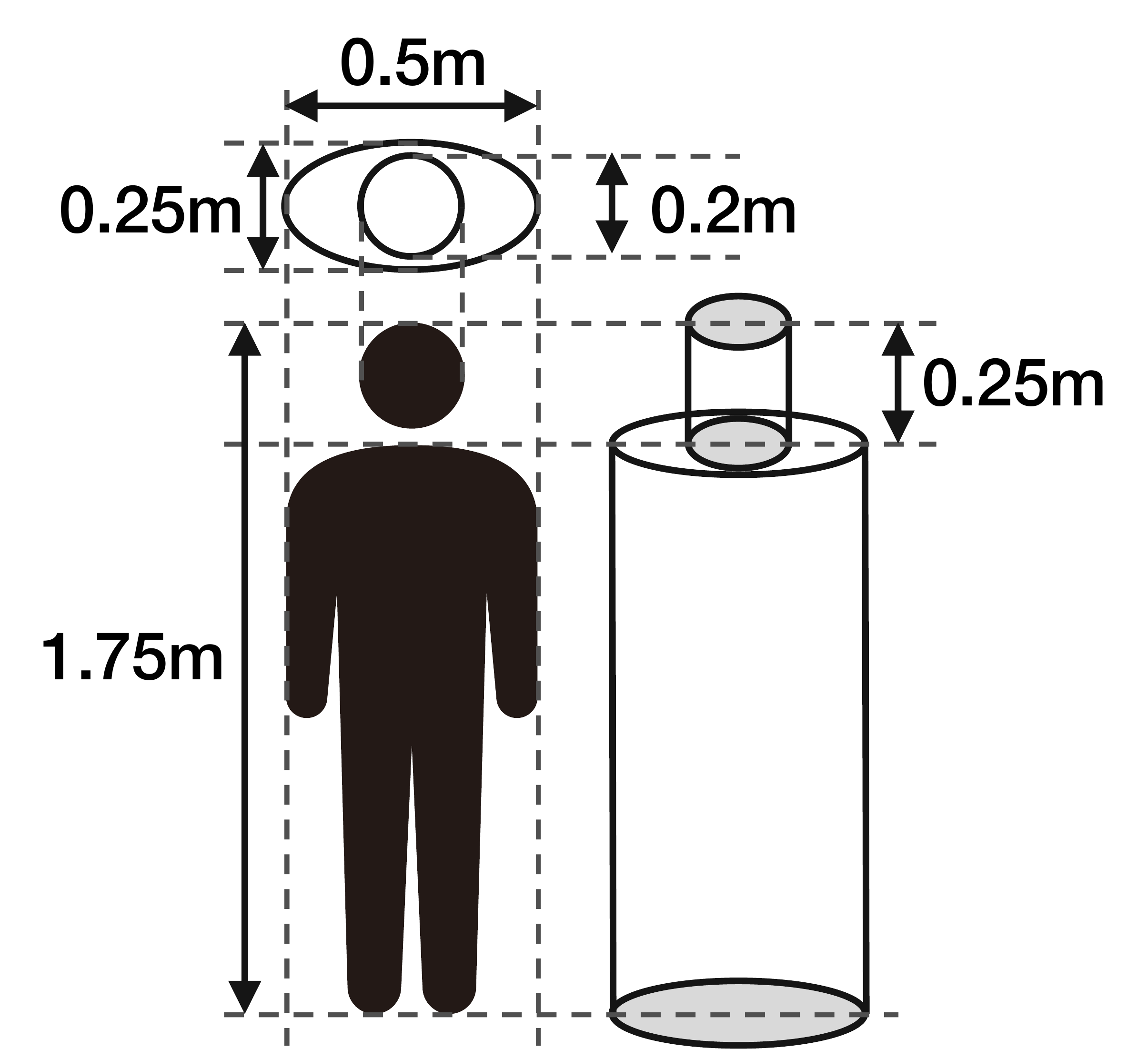}
  \caption{Human model. We use a twin cylinder model \cite{twin_cylinder} as the human shape model.}
  \label{fig:twin_cylinder}
\end{figure}

\begin{figure*}[!t]
    \centering
    \subfloat[$\mathrm{A_{low}}$ \label{fig:camera_image_a1}]{
        \includegraphics[width=0.2\textwidth]{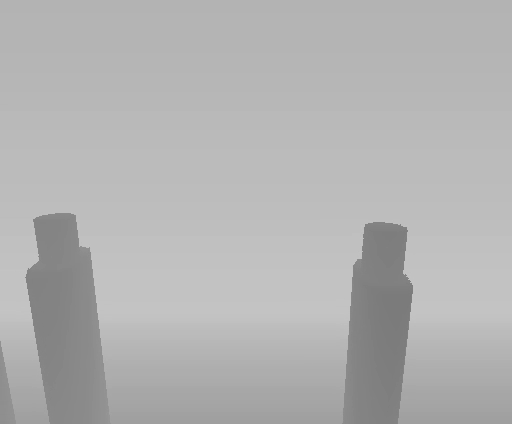}
    }
    \subfloat[$\mathrm{B_{low}}$ \label{fig:camera_image_b1}]{
        \includegraphics[width=0.2\textwidth]{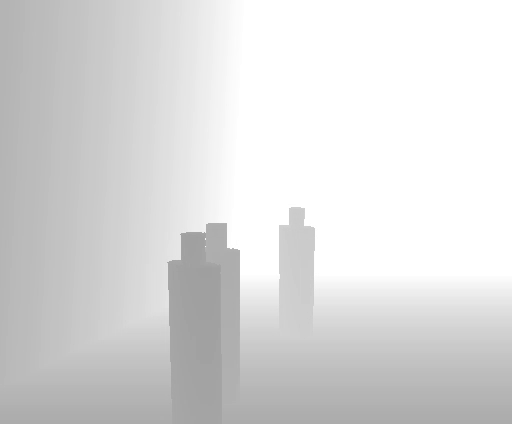}
    }
    \subfloat[$\mathrm{C_{low}}$ \label{fig:camera_image_c1}]{
        \includegraphics[width=0.2\textwidth]{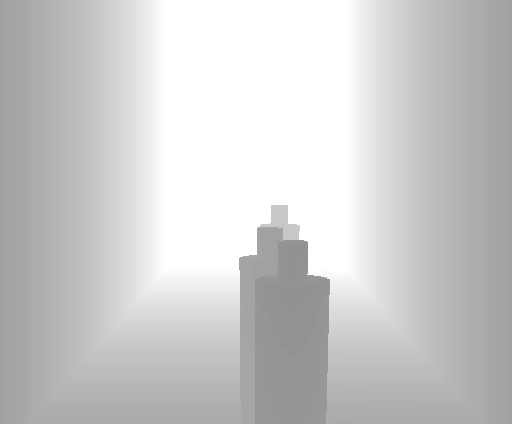}
    }
    \subfloat[$\mathrm{D_{low}}$ \label{fig:camera_image_d1}]{
        \includegraphics[width=0.2\textwidth]{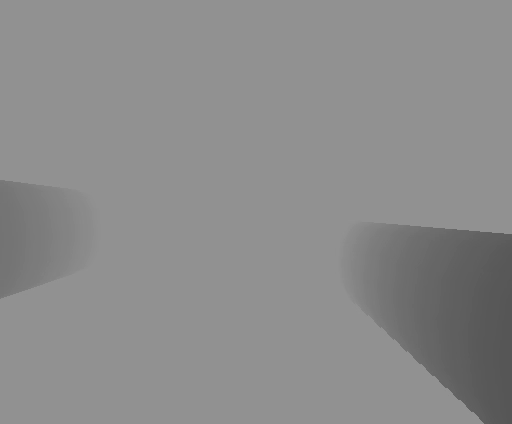}
    }

    \subfloat[$\mathrm{A_{high}}$ \label{fig:camera_image_a2}]{
        \includegraphics[width=0.2\textwidth]{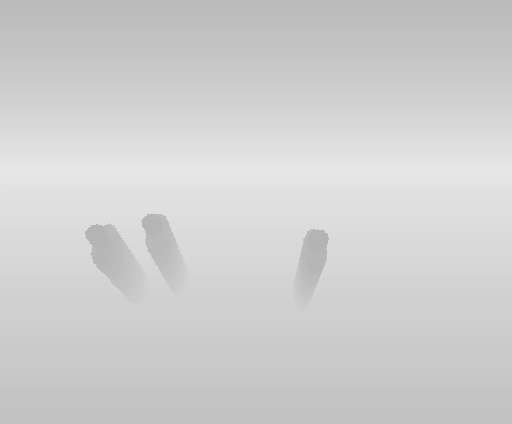}
    }
    \subfloat[$\mathrm{B_{high}}$ \label{fig:camera_image_b2}]{
        \includegraphics[width=0.2\textwidth]{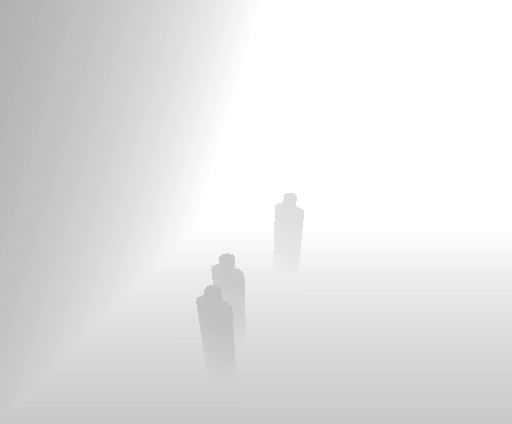}
    }
    \subfloat[$\mathrm{C_{high}}$ \label{fig:camera_image_c2}]{
        \includegraphics[width=0.2\textwidth]{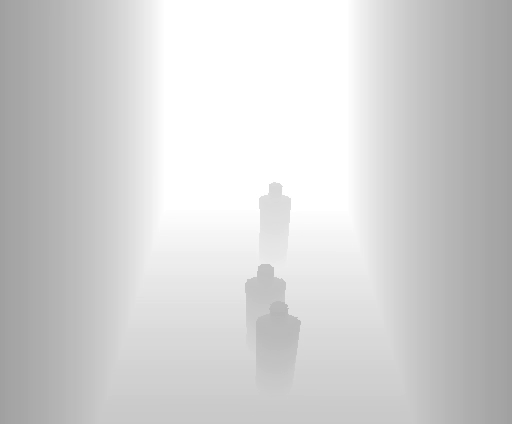}
    }
    \subfloat[$\mathrm{D_{high}}$ \label{fig:camera_image_d2}]{
        \includegraphics[width=0.2\textwidth]{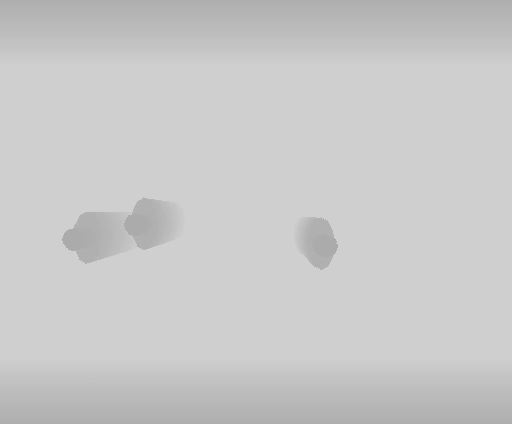}
    }
    \caption{Examples of camera images. The camera positions are referred to in Fig. \ref{fig:passage}. \label{fig:camera_image}}
\end{figure*}

A 3D spatial model of the mmWave communication area is generated first.
Then, using the 3D spatial model and mmWave propagation model, the received power were calculated using a radio propagation simulator.
Simultaneously, using the 3D spatial model and 3D CG techniques, camera images were captured by a 3D CG modeler.
Using this generated dataset of the received power and images, we evaluated the prediction accuracy of each algorithm.

We consider the 3D passage model with a point-to-point mmWave communication system presented in Figs. \ref{fig:camera_image_blender} and \ref{fig:passage}.
The length and width of the passage were 10\,m and 4\,m, respectively, and the height of the walls was 10\,m.
An AP, an STA, and a camera were present in the passage.
Letting the center of the passage be the coordinate origin $\mathrm{O}$, the AP was placed at $(x,y,z) = (0, -2, 2.25)$ and the STA was placed at $(0, 2, 1.25)$.
The units of the $x$, $y$, and $z$ axes were meters.
We considered that the AP and STA were constantly communicating with each other using a radio frequency of $60$\,GHz.
The AP antenna was a directional antenna with a 15-degree beam width and $24$\,dBi gain at $60$\,GHz.
The antenna was facing toward the STA and the transmission power was $100$\,mW.
The STA antenna was an omnidirectional antenna and the minimum sensitivity was set to $-68$\,dBm, which is the required sensitivity for the lowest PHY rate for data transmission in IEEE 802.11ad \cite{11ad}.
As a result, the received power available for learning was greater than or equal to $-68$\,dBm, thus the predicted received power was also approximately greater than or equal to $-68$\,dBm.
The camera was placed at one of the eight positions shown in Fig.~\ref{fig:passage}.

The pedestrians came in from either side of the passage and walked straight, as shown in Fig.~\ref{fig:passage}, with speeds in the range of $[0.5,2]$\,m/s.
The y-directional positions of the pedestrians who walked from the left side to the right side were distributed uniformly in the range $[-1.75,0]$\,m and those of pedestrians who walked from the right side to the left side were distributed uniformly in the range $[0,1.75]$\,m.
We set pedestrians entering the passage to follow a Poisson distribution with $\lambda = 0.25$ \cite{pedestrian_poisson}.
Since we primarily concerned with the situation only when some pedestrians block the LOS path simultaneously, we considered Poisson arrivals to be an adequate scenario for our study.

The shape of a pedestrian was modeled by a twin cylinder model \cite{twin_cylinder}, as shown in Fig.~\ref{fig:twin_cylinder}. This is a model for a calculation of the human body shadowing in the mmWave communications.
The magnetic permeability and electric constants of human bodies were assumed to be the same values as those of water. This provided shadowing properties similar to those of real humans \cite{human_phantoms}.
We calculated the received power using the radio propagation simulator RapLab \cite{RapLab}, which conducts 3D ray tracing using imaging methods.

We employed a depth camera model equivalent to the Kinect v2, which has a resolution of $512 \times 424$ pixels, an angular field of view of $60^\circ$ in the vertical plane, $70^\circ$ in the horizontal plane, a frame rate of 30\,fps, and a depth capture range of 0.5--8\,m.
We considered eight patterns for the camera position of A: $(0,-2)$, B: $(-4,-2)$, C: $(-4,0)$, or D: $(0,0)$ in the X-Y plane and a height of 2.25\,m (low) or 5\,m (high). This was in order to evaluate the accuracy of the proposed prediction scheme with respect to camera positions.
The camera at each position was pointed at $(0,0,1.75)$, which was the midpoint between the AP and the STA.
We generated the depth images using the 3DCG modeler Blender \cite{blender}.
The example images from various camera positions are shown in Fig. \ref{fig:camera_image}.

The algorithms predicted a received power in 0\,ms (current) or 500\,ms (future) ahead. Therefore, $k$ was set to $0$ and $15$, respectively.
In the channel model for 60\,GHz WLAN systems \cite{11ad_channel_model}, the duration of the signal level degradation is characterized by a Weibull distribution with the shape parameter $\alpha = 6.32$ and the scale parameter $\beta = 0.59$.
Hence, the average duration of the degradation is 0.55\,s, so that if we predicted a received power 500\,ms ahead, we could get most of the received power time variation during the blockage.

The simulation ran for 30\,min and generated a dataset with $54000$ samples of the received power and images.
Thus, $43200$ samples were used for training, and $10800$ samples were used for the performance evaluation.

\subsection{Prediction Results for Simulation Data}
\label{subsec:simulation_result}

\begin{figure*}[!t]
  \centering
  \subfloat[When the camera was at $\mathrm{A_{low}}$.\label{fig:result_time_simu_a1}]{
      \includegraphics[width=0.45\textwidth]{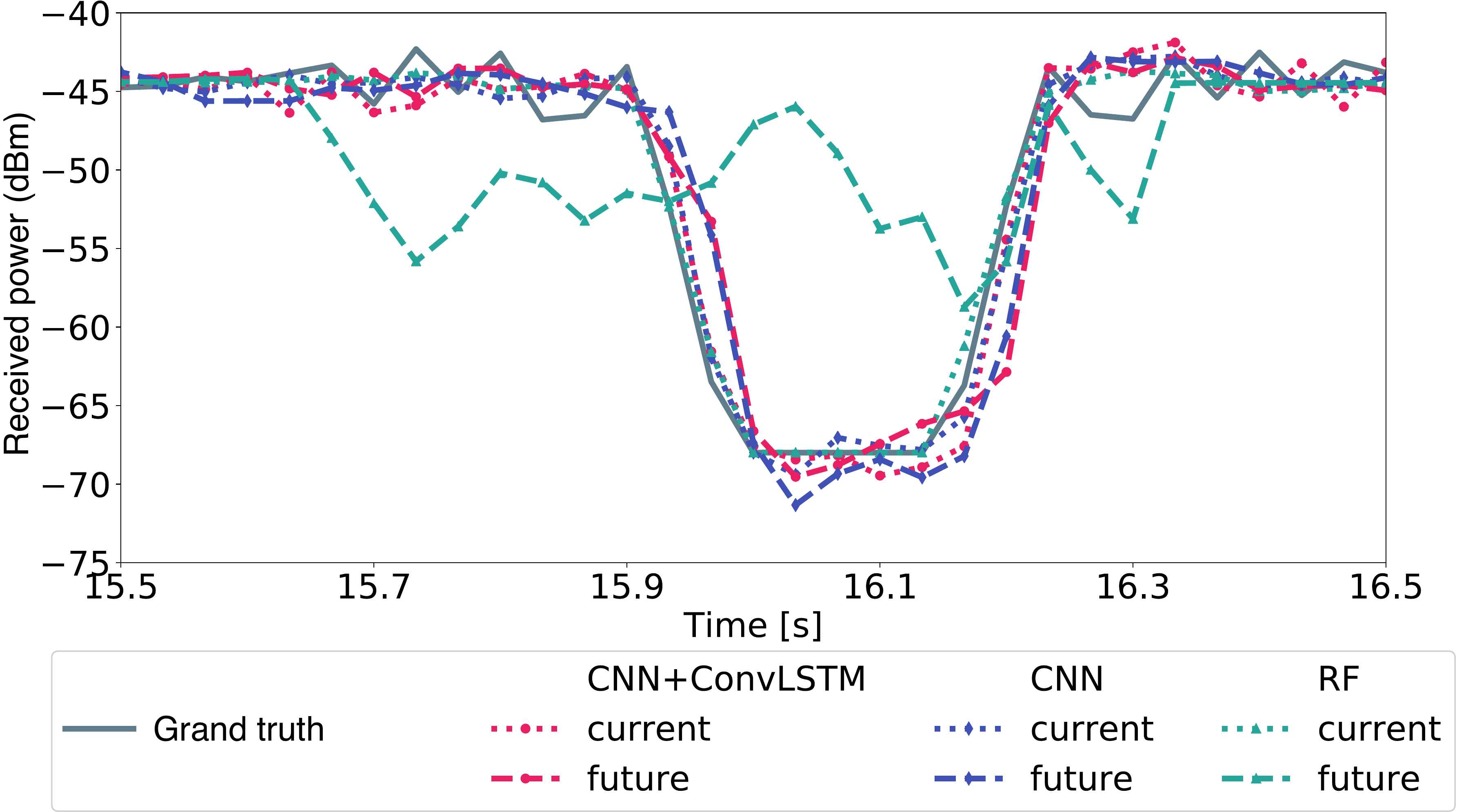}
  }\hspace{0.3em}
  \subfloat[When the camera was at $\mathrm{A_{high}}$.\label{fig:result_time_simu_a2}]{
      \includegraphics[width=0.45\textwidth]{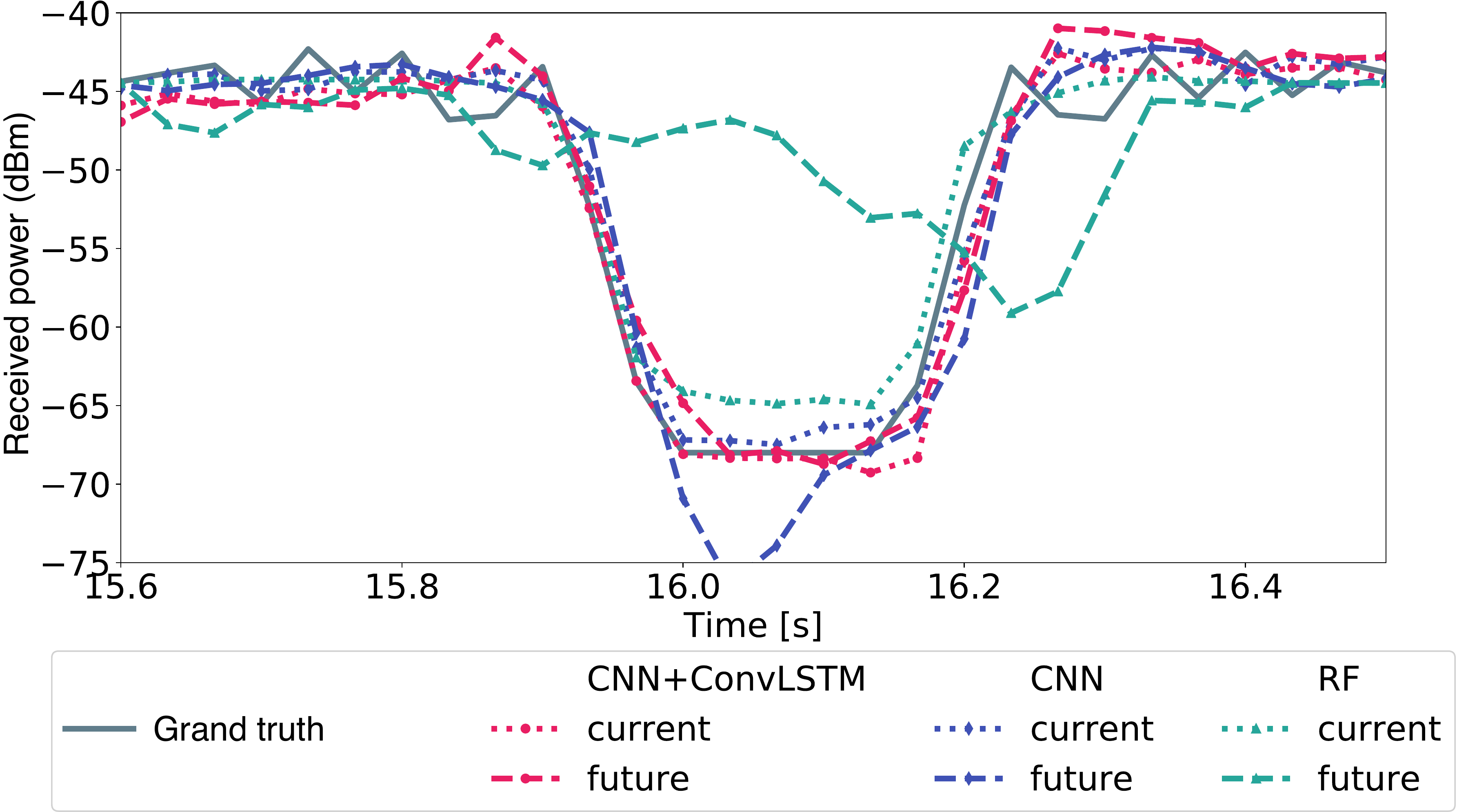}
  }\\
  \subfloat[When the camera was at $\mathrm{C_{low}}$.\label{fig:result_time_simu_c1}]{
      \includegraphics[width=0.45\textwidth]{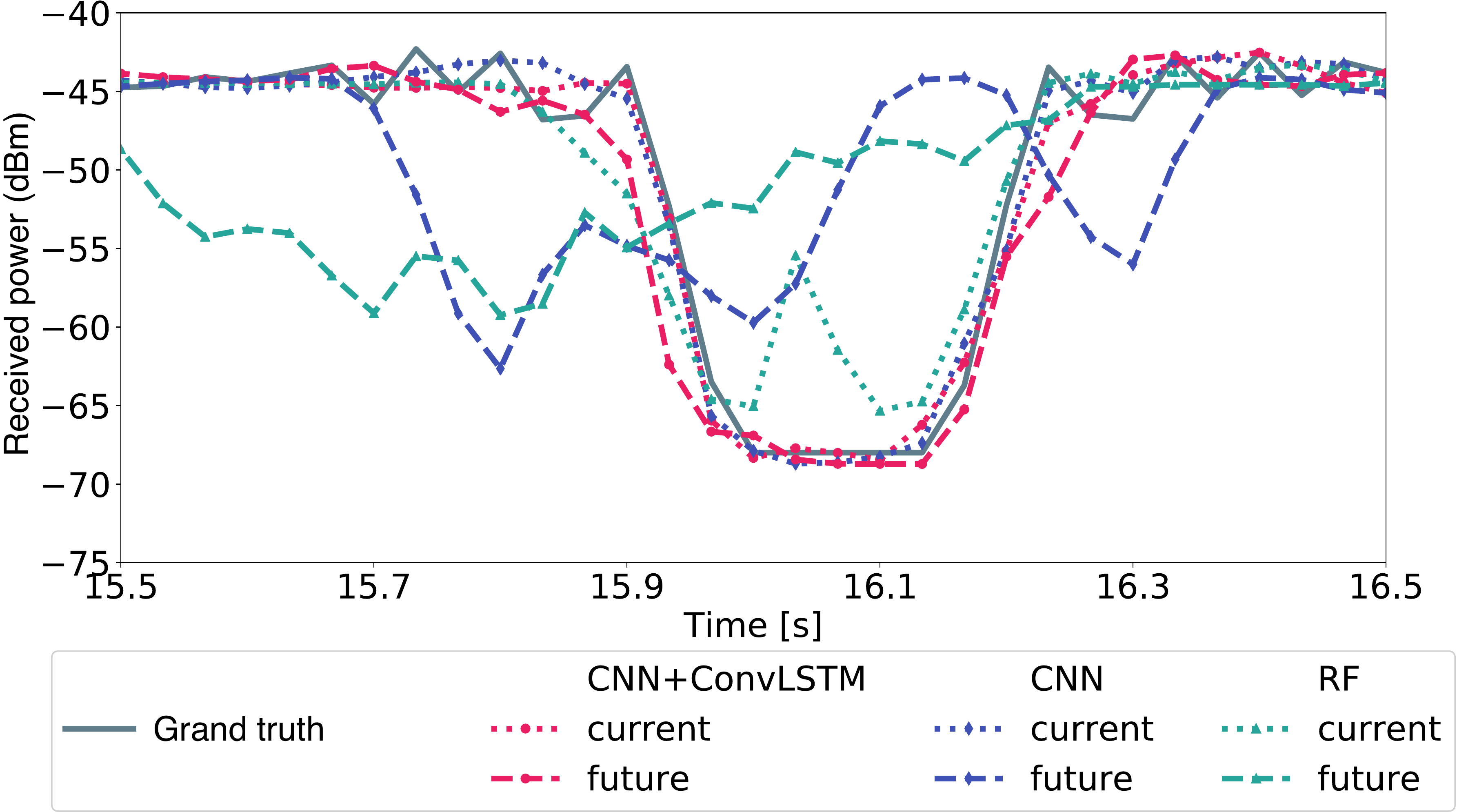}
  }\hspace{0.3em}
  \subfloat[When the camera was at $\mathrm{C_{high}}$.\label{fig:result_time_simu_c2}]{
      \includegraphics[width=0.45\textwidth]{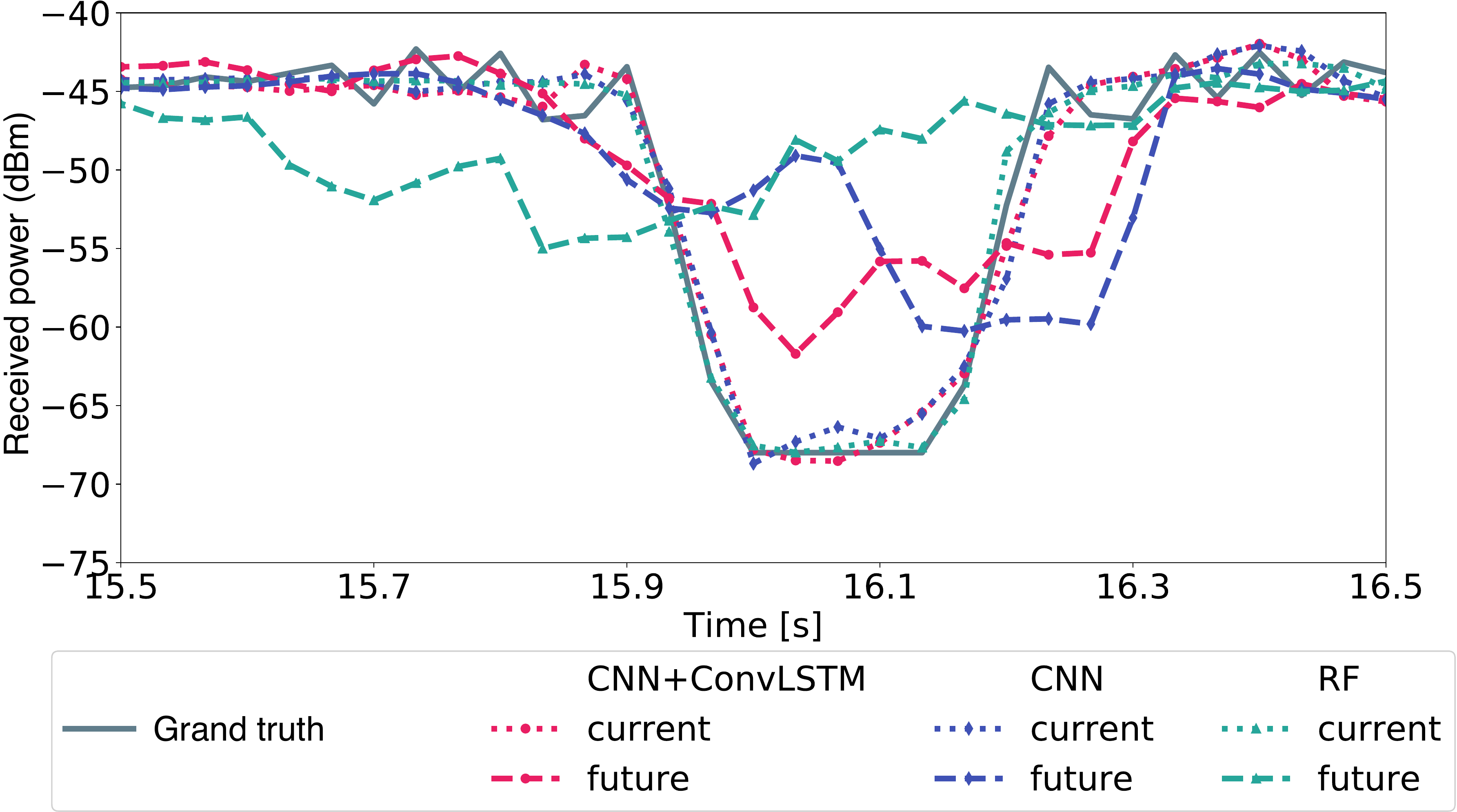}
  }\\
  \subfloat[Another portion of the received power time series when the camera was at $\mathrm{C_{low}}$. In this portion, pedestrians blocked the view of the camera.\label{fig:result_time_simu_c1_2}]{
      \includegraphics[width=0.45\textwidth]{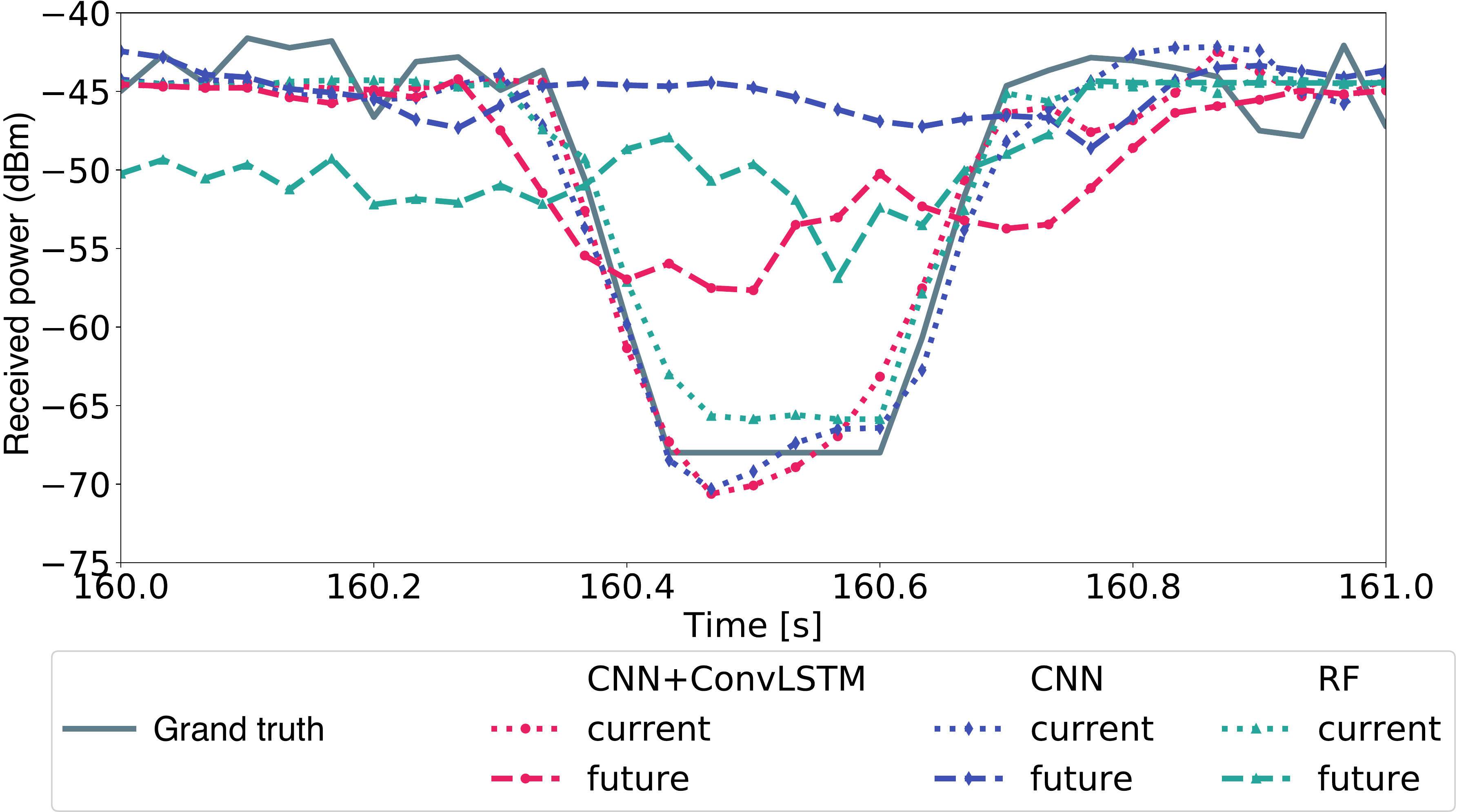}
  }\hspace{0.3em}
  \subfloat[Another portion of the received power time series when the camera was at $\mathrm{C_{high}}$. The view of the camera was less likely to be blocked owing to its higher position than the camera at $\mathrm{C_{low}}$. \label{fig:result_time_simu_c2_2}]{
      \includegraphics[width=0.45\textwidth]{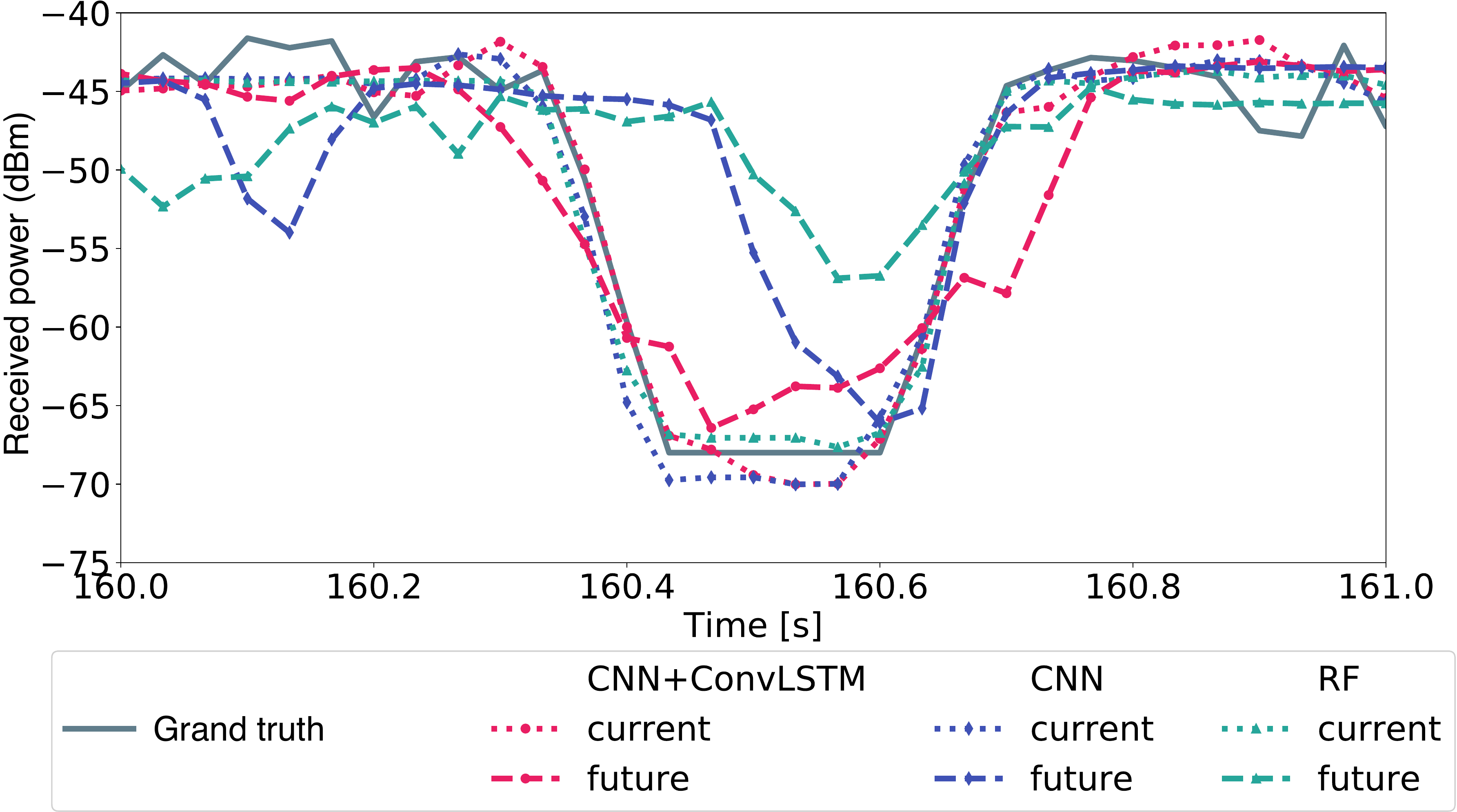}
  }
  \caption{Grand truth (simulated received power) and predicted received power as a function of elapsed time. \label{fig:result_time_simu}}
\end{figure*}

\begin{figure}[!t]
    \centering
    \subfloat[When the current received power (0\,ms ahead) is predicted.\label{fig:result_sim_0}]{
        \includegraphics[width=0.45\textwidth]{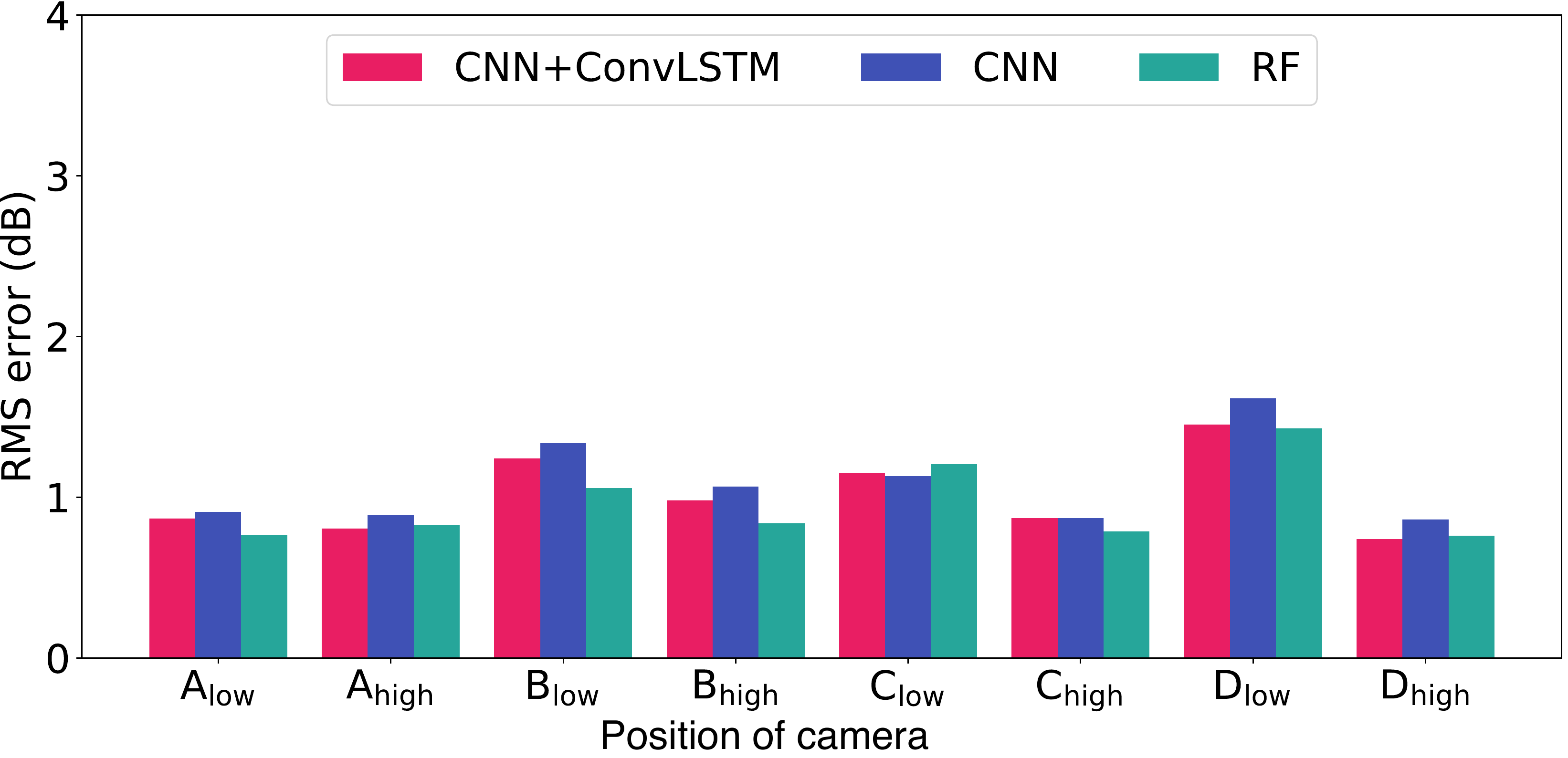}
    }\\
    \subfloat[When the future received power (500\,ms ahead) are predicted.\label{fig:result_sim_15}]{
        \includegraphics[width=0.45\textwidth]{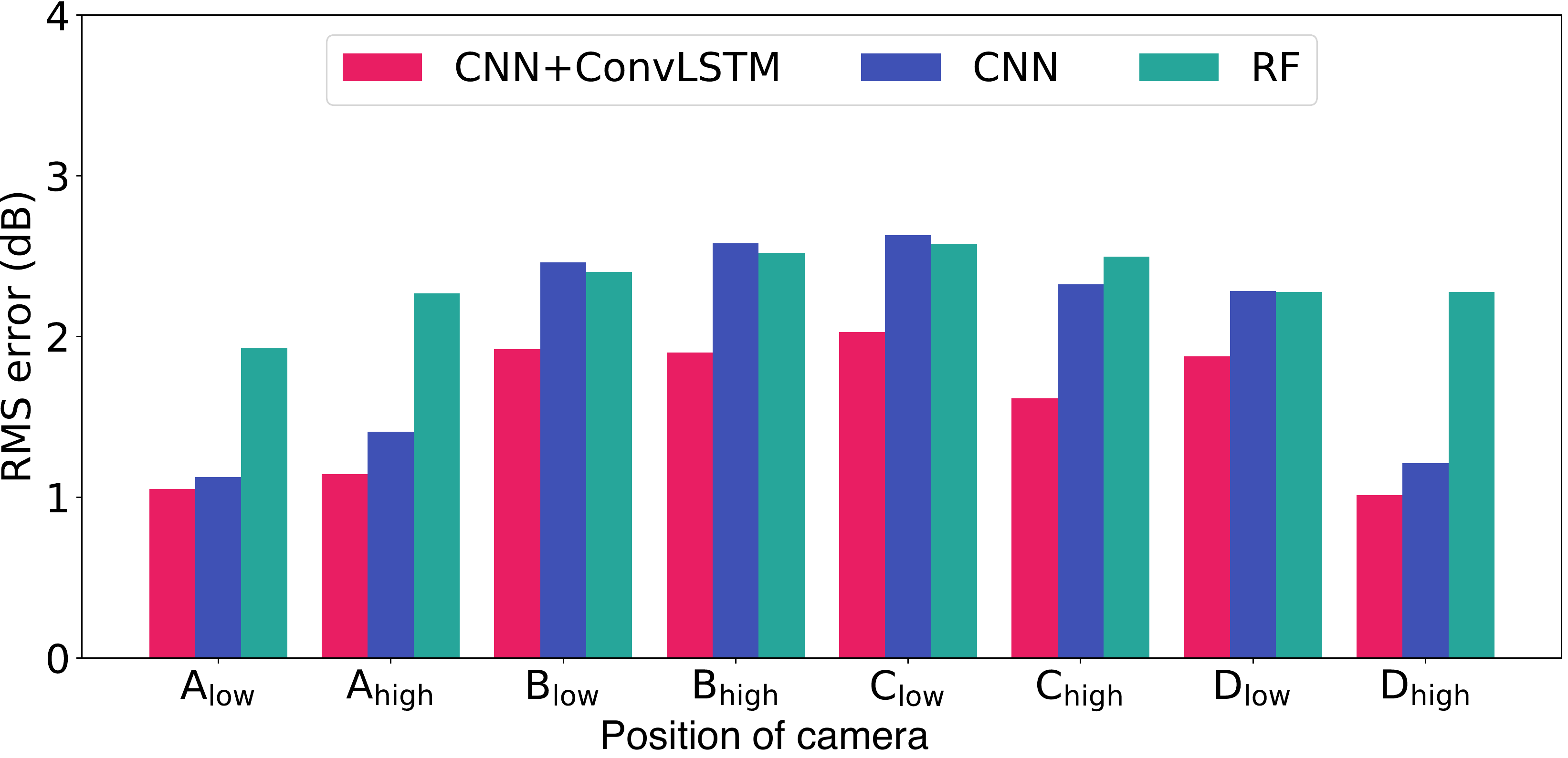}
    }
    \caption{RMS error of each algorithm. \label{fig:result_rmse_sim}}
  \end{figure}

Fig.~\ref{fig:result_time_simu} presents the time series of the simulated received power and the received power predicted by each algorithm when the camera was placed at position $\mathrm{A_{low}}$, $\mathrm{A_{high}}$, $\mathrm{C_{low}}$, and $\mathrm{C_{high}}$.
The prediction results of the current received power which are indicated by thin dotted lines in Fig.~\ref{fig:result_time_simu} match the grand truth (the simulated received power).
Fig.~\ref{fig:result_rmse_sim}\subref{fig:result_sim_0} shows the RMS errors for the current power prediction in each camera position. The RMS errors in the camera positions except $\mathrm{B_{low}}$, $\mathrm{C_{low}}$, and $\mathrm{D_{low}}$ achieved even less than 1\,dB.

The prediction results of the future received power which are indicated by thick dotted lines in Fig.~\ref{fig:result_time_simu} also match the grand truth except the results of the RF.
In particular, the CNN+ConvLSTM predicted the future received power with the lowest RMS error among the models for all camera positions as shown in Fig.~\ref{fig:result_rmse_sim}. As mentioned in Section \ref{sec:ml_algorithm_and_model}, the ConvLSTM employs both convolutional and recurrent architectures to capture spatiotemporal features, and we consider that the architectures contribute to the accurate prediction. 
The RF predicted the received power with larger errors than that of the CNN+ConvLSTM since the RF has less capability to model the spatiotemporal features than ConvLSTM.
The CNN predicted the future received power with small errors when the camera was at $\mathrm{A}$ and $\mathrm{D_\mathrm{high}}$, but the large RMS errors occurred for the other camera positions. In CNN+ConvLSTM, the RMS errors also increased for the camera positions $\mathrm{B}$ and $\mathrm{C}$.
One of the reasons why the error increased is the existence of blind spots, i.e., the state of the whole LOS path of mmWave link was not observed in the images. When the camera was placed at $\mathrm{D_\mathrm{low}}$ a part of the LOS path was not included in the FOV of the camera. Therefore, the pedestrians in the blind spots cannot be recognized. When the camera was placed at $\mathrm{B}$ and $\mathrm{C}$ the pedestrian can block the FOV of the camera like in Figs.~\ref{fig:camera_image}\subref{fig:camera_image_b1} and \ref{fig:camera_image}\subref{fig:camera_image_c1}. The FOV blockage caused the blind spots temporally and the large prediction errors were likely to occur.
The FOV of the camera could also be blocked by pedestrians when the camera was placed at $\mathrm{A}$. However, when the LOS path was not observed in the FOV of the camera due to the blockage, the LOS path was also simultaneously blocked, i.e., whether the LOS path was blocked or not depended on whether the entire LOS path was observed or not.
Hence, the NNs could accurately predict the received power even when the view of the camera was being blocked at camera position $\mathrm{A}$.

Higher camera positions made the camera view less likely to be blocked by pedestrians and could prevent the degradation of the received power prediction accuracy caused by blockages of the camera view as shown in Figs.~\ref{fig:result_time_simu}\subref{fig:result_time_simu_c1_2} and \ref{fig:result_time_simu}\subref{fig:result_time_simu_c2_2}.
On the other hand, the higher camera positions made the pedestrians smaller in the images and made it more difficult to predict the movement of the pedestrians.
Hence, the higher camera positions could degrade the prediction accuracy, as shown in Figs.~\ref{fig:result_time_simu}\subref{fig:result_time_simu_c1} and \ref{fig:result_time_simu}\subref{fig:result_time_simu_c2}. This was especially true when a few pedestrians were in the passage and prediction accuracy was already high for a lower camera position.

In future received power predictions, the camera position at $\mathrm{A}$ and $\mathrm{D_{high}}$ enabled the most-accurate prediction of the received power for all algorithms.
These results suggest that a camera position over the AP or a position overlooking the passage provides a higher prediction accuracy.


\begin{figure}[!t]
    \centering
    \includegraphics[width=0.4\textwidth]{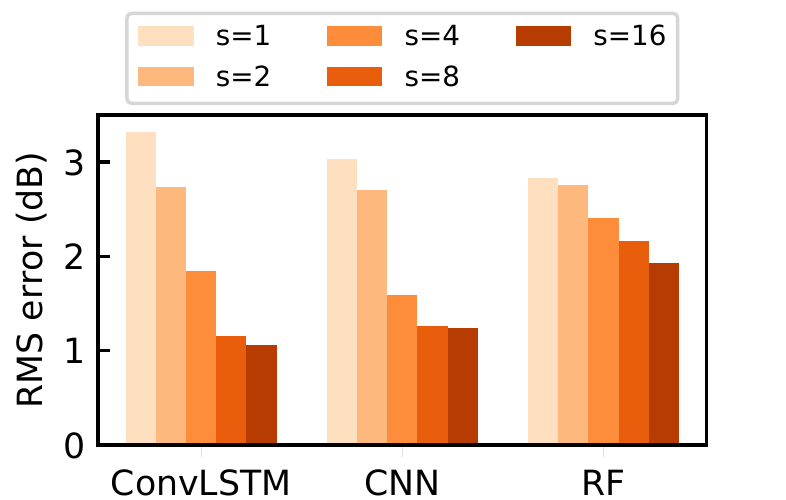}
    \caption{RMS errors as a function of the number of images in a time-sequential image, $s$. The camera is located at $\mathrm{A}_\mathrm{low}$.}
    \label{fig:ts}
\end{figure}

Fig.~\ref{fig:ts} shows the RMS errors as a function of the number of images in a time-sequential image, $s$. As increasing $s$, the RMS errors decrease for all the models. Especially, the time-sequential image works well in CNN+ConvLSTM and CNN. This is because that the large $s$ can provide the longer-term temporal features for the models and the ConvLSTM and 3D Conv layers captured the features more effectively. However, the larger $s$ makes the ML model bigger and increases the computation time for training. In our ML experiments, the average computation time for training increased almost linearly as $s$ increased. Training the CNN+ConvLSTM model took around 16 hours when $s=16$, while the RF took only ten to twenty minutes. Moreover, the larger model consumes the larger memory capacity. In our experimental setup, $s$ larger than 16 was difficult to train due to the limitations of the memory capacity and the computation time. Thus, we used $s=16$.
These results suggest that we should carefully select the ML model and the hyperparameters so as to meet the system or application requirements.

On the other hand, the average computation time for predicting a received power from time-sequential images was much shorter than the required time for training.
The time were about 2.9\,ms, 2.1\,ms, and 0.20\,ms for the CNN+ConvLSTM, 3D CNN, and RF, respectively.
These periods were sufficiently shorter than the interval of image acquisition from the camera, the beacon interval of the WLANs, and the superframe interval of the current LTE networks. Therefore, it is expected that the proposed prediction mechanism works in a real-time basis.

\section{Evaluation Using Experimental Data}
\label{sec:experiments}
We evaluated the prediction accuracy of the proposed mechanism by using experimental data obtained from experiments using IEEE 802.11ad devices and an RGB-D camera Kinect. 
The same experimental setup for ML as that in Section \ref{sec:simulation} is used for the received power prediction.

\begin{figure}[!t]
  \centering
  \includegraphics[width=0.35\textwidth]{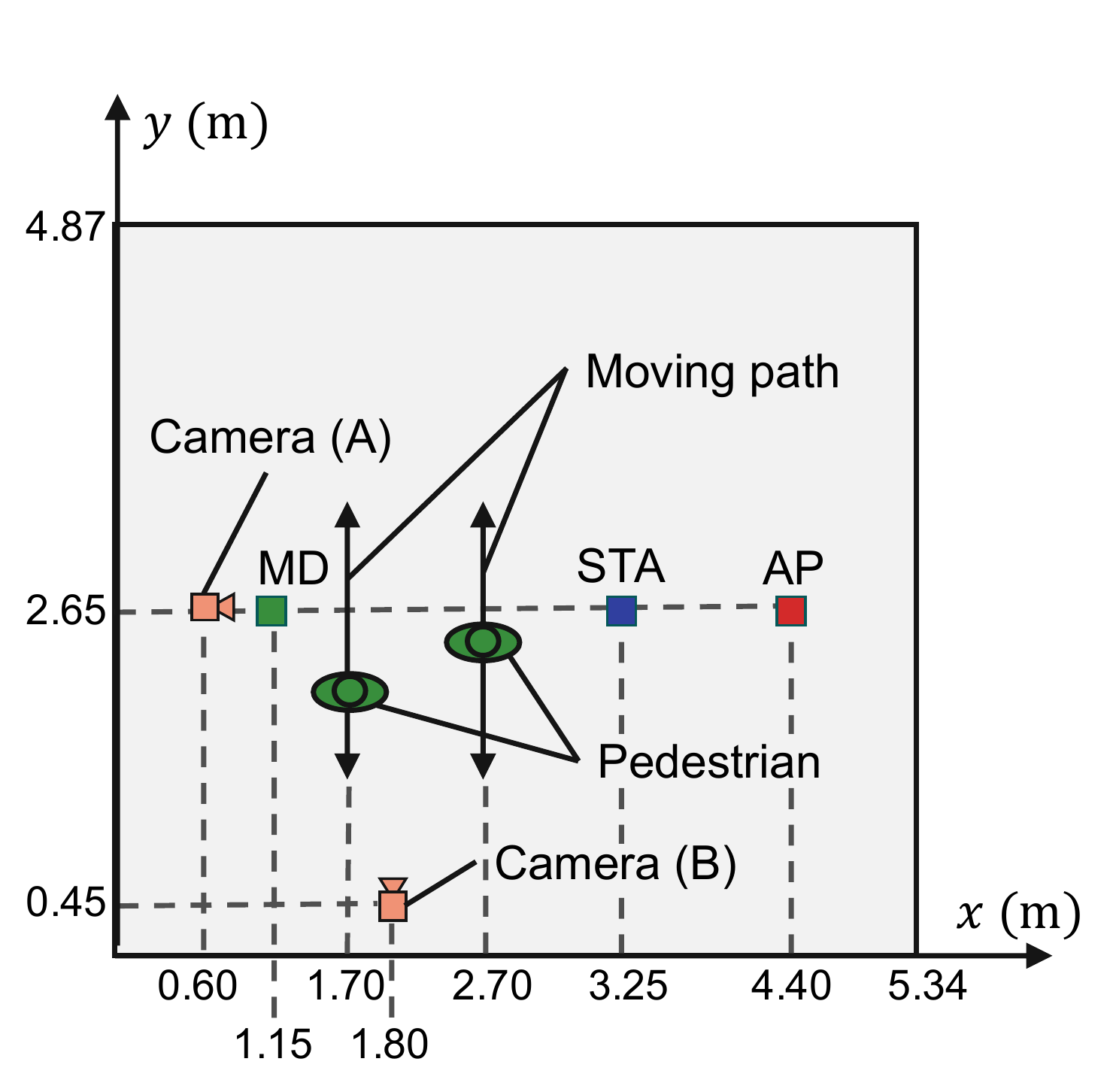}
  \caption{Top view of experimental environment. The measurement device (MD) captures the received power of the signals transmitted from the AP to the STA.}
  \label{fig:room}
\end{figure}

\begin{figure}[!t]
  \centering
  \subfloat[At position A.\label{fig:ss_exp_a}]{
    \fbox{\includegraphics[width=0.2\textwidth]{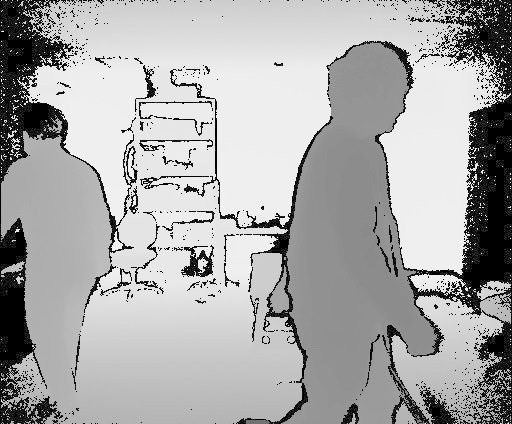}}
  }\hspace{1em}
  \subfloat[At position B.\label{fig:ss_exp_b}]{
      \fbox{\includegraphics[width=0.2\textwidth]{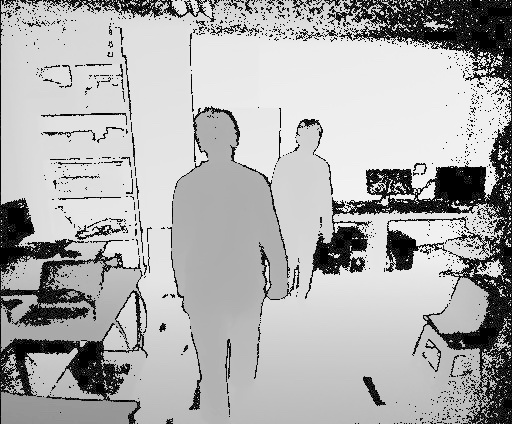}}
  }
  \caption{Example depth images at each camera position. The camera positions are referred to in Fig. \ref{fig:room}. \label{fig:ss_exp}}
\end{figure}

\begin{table}[!t]
\centering
\caption{Experimental equipment.}
\label{tab:equipments}
\begin{tabular}{cc}
\toprule
IEEE 802.11ad WLAN AP   & Dell Wireless Dock D5000 \\
IEEE 802.11ad WLAN STA  & Dell Latitude E5540      \\
Measurement device antenna     & Horn antenna, 24\,dBi \\
\multirow{2}{*}{RGB-D camera}  & Microsoft Kinect for Windows \\[-1mm]
                               & (Model:1656) \\
\bottomrule
\end{tabular}
\end{table}

\subsection{Setup of mmWave Experiments}
To measure a time-varying received power of IEEE 802.11ad WLAN signals in the 60\,GHz band, we developed a received power measurement system proposed in a previous work \cite{koda18}.
The system consisted of an AP, an STA, a measurement device (MD) and an \mbox{RGB-D} camera as shown in Fig.~\ref{fig:room}.
The received power measured at the MD was used for the prediction. That is, the MD was regarded as a practical STA in the experiments.
The AP, the STA, and the MD were placed at the heights of 0.70\,m, 0.65\,m, and 0.85\,m, respectively.
The camera was placed in either position A or B in Fig.~\ref{fig:room}, and images from the camera positions are shown in Fig.~\ref{fig:ss_exp}.
The height of position A was 1.50\,m and that of B was 1.25\,m.
Details of the experimental equipment are provided in TABLE~\ref{tab:equipments}.
The AP, the STA, and the RGB-D camera were all commercial products.

IEEE 802.11ad devices perform beamforming autonomously when a link quality is degraded, and the proposed mechanism can learn from the received power that resulted from the beamforming. However, since the implementation of beamforming operations depends on manufacturers and it is black-boxed, effects of the black-boxed beamforming operations can reduce the reproducibility of the experiments. 
Therefore, we set up the experiment so that the beamforming is not performed even when the blockage occurs.
Specifically, the MD was located behind the STA, two pedestrians traveled between the STA and the MD moving along the path shown in Fig.~\ref{fig:room}.
This prevented the AP and STA from performing the beamforming because the received power at the AP and STA was not changed even when the LOS path between the STA and MD was blocked. 
Thus the measured received power values at the MD was not affected by the beamforming control.
The MD captured the 60.48\,GHz signal and converted it to an intermediate-frequency (IF) at 2.98\,GHz. Then, a spectrum analyzer, equipped with the MD, measured the power of the IF signal.
We treated this measured signal power as the received power at the MD.
Although the measurement was limited by the noise floor of the spectrum analyzer, the received power greater than approximately $-40$\,dBm could be measured.

Each pedestrian moved with a random speed and blocked the LOS path almost every 6\,s.
The experiment ran for approximately 10\,min and models were trained for 100 epochs.
The other hyperparameters were the same as for the simulation.

\subsection{Experimental Results and Evaluation}

\begin{figure*}[!t]
  \centering
  \subfloat[Current received power prediction when camera was at A.\label{fig:result_time_exp_a_current}]{
      \includegraphics[width=0.45\textwidth]{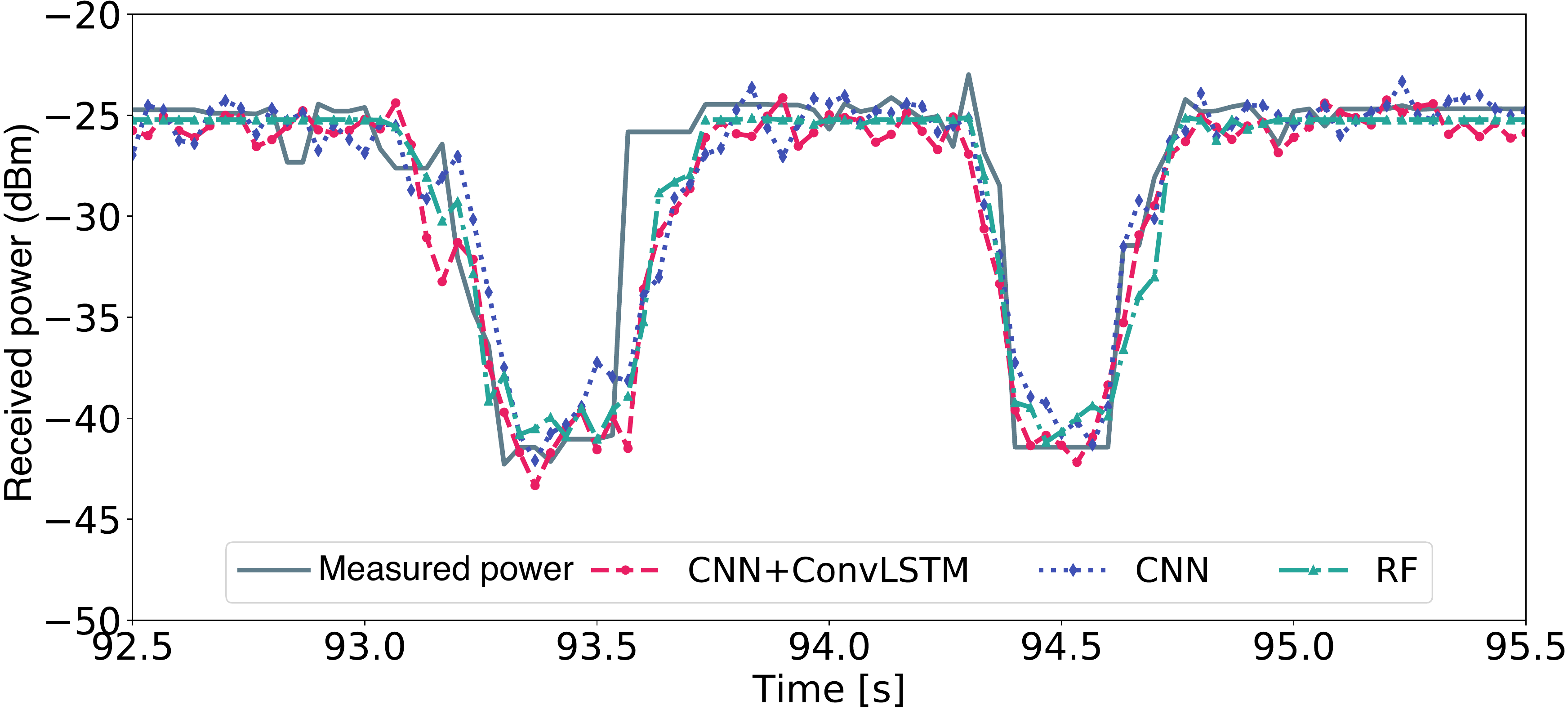}
  }\hspace{0.3em}
  \subfloat[Future received power prediction (500\,ms ahead) when camera was at A.\label{fig:result_time_exp_a_future}]{
      \includegraphics[width=0.45\textwidth]{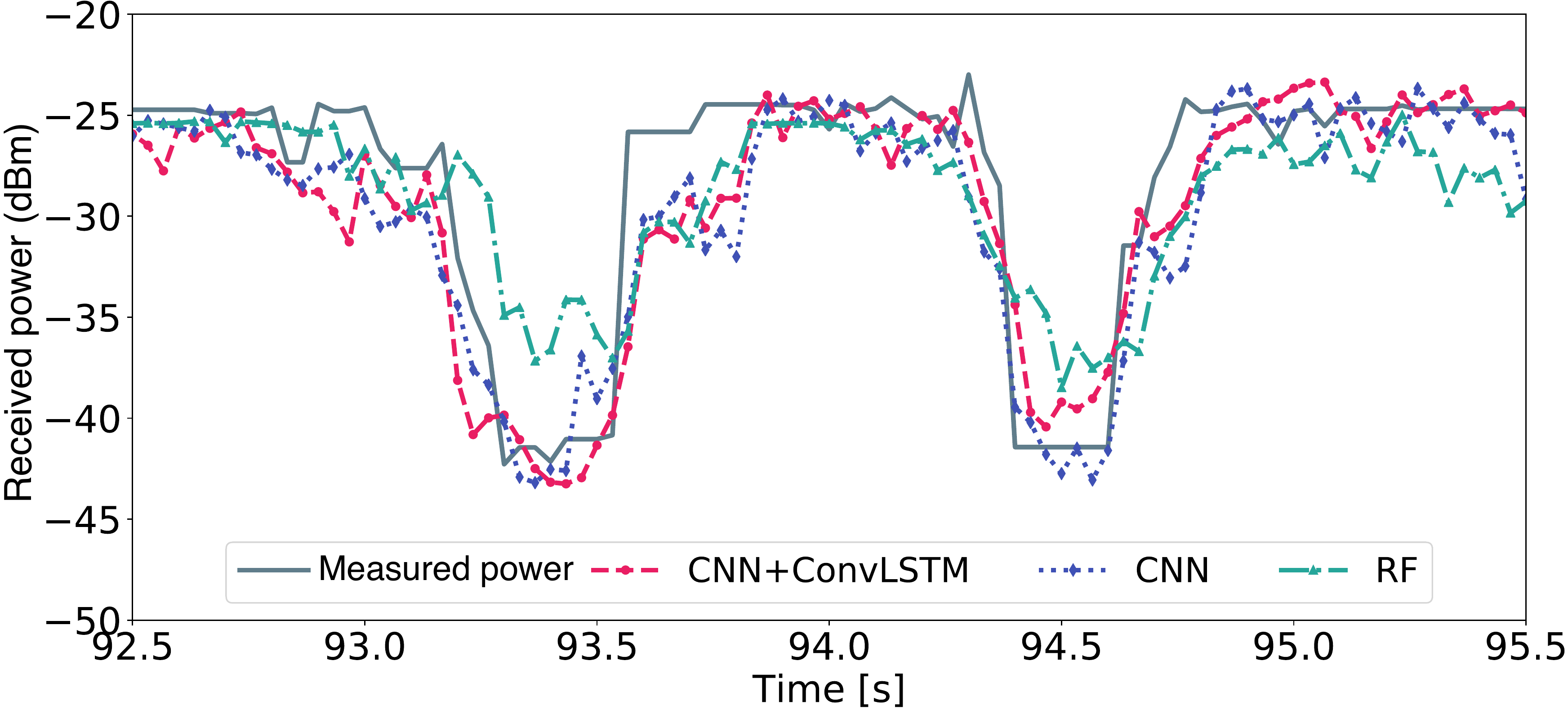}
  }
  \\
  \subfloat[Current received power prediction when camera was at B.\label{fig:result_time_exp_b_current}]{
      \includegraphics[width=0.45\textwidth]{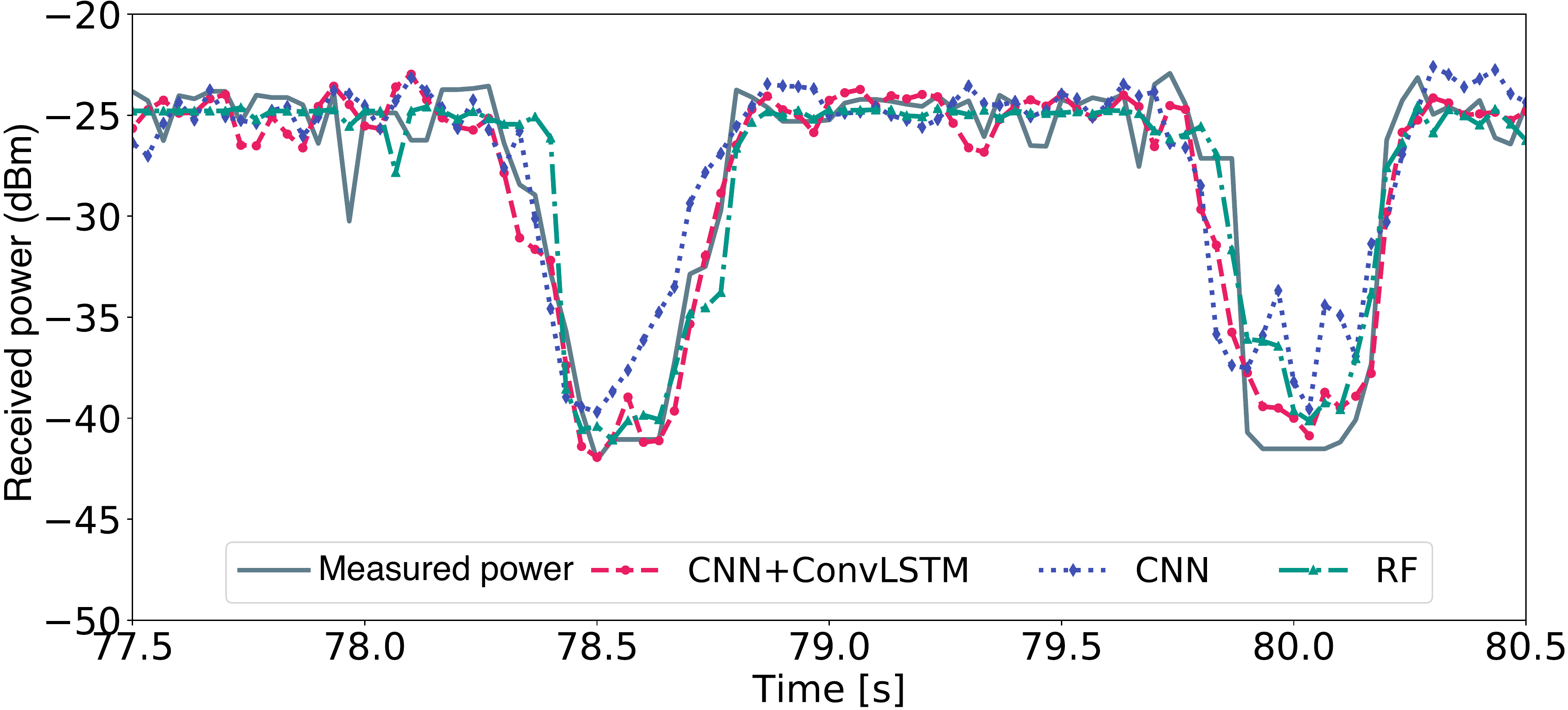}
  }\hspace{0.5em}
  \subfloat[Future received power prediction (500\,ms ahead) when camera was at B.\label{fig:result_time_exp_b_future}]{
      \includegraphics[width=0.45\textwidth]{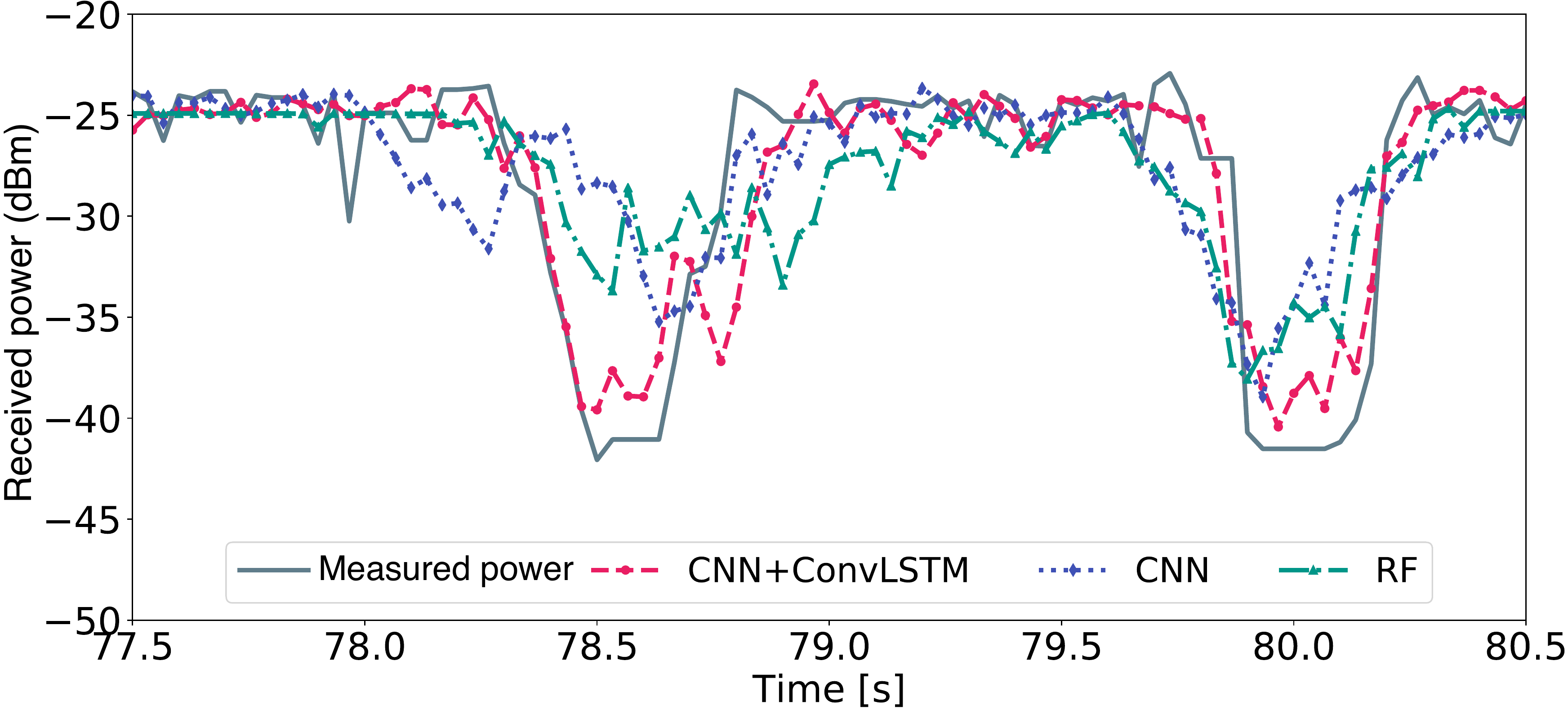}
  }
  \\
  \subfloat[Another portion of the predicted current received power when the camera was at A. In this portion, pedestrians simultaneously blocked the LOS path.\label{fig:result_time_exp_a_dbl_current}]{
      \includegraphics[width=0.45\textwidth]{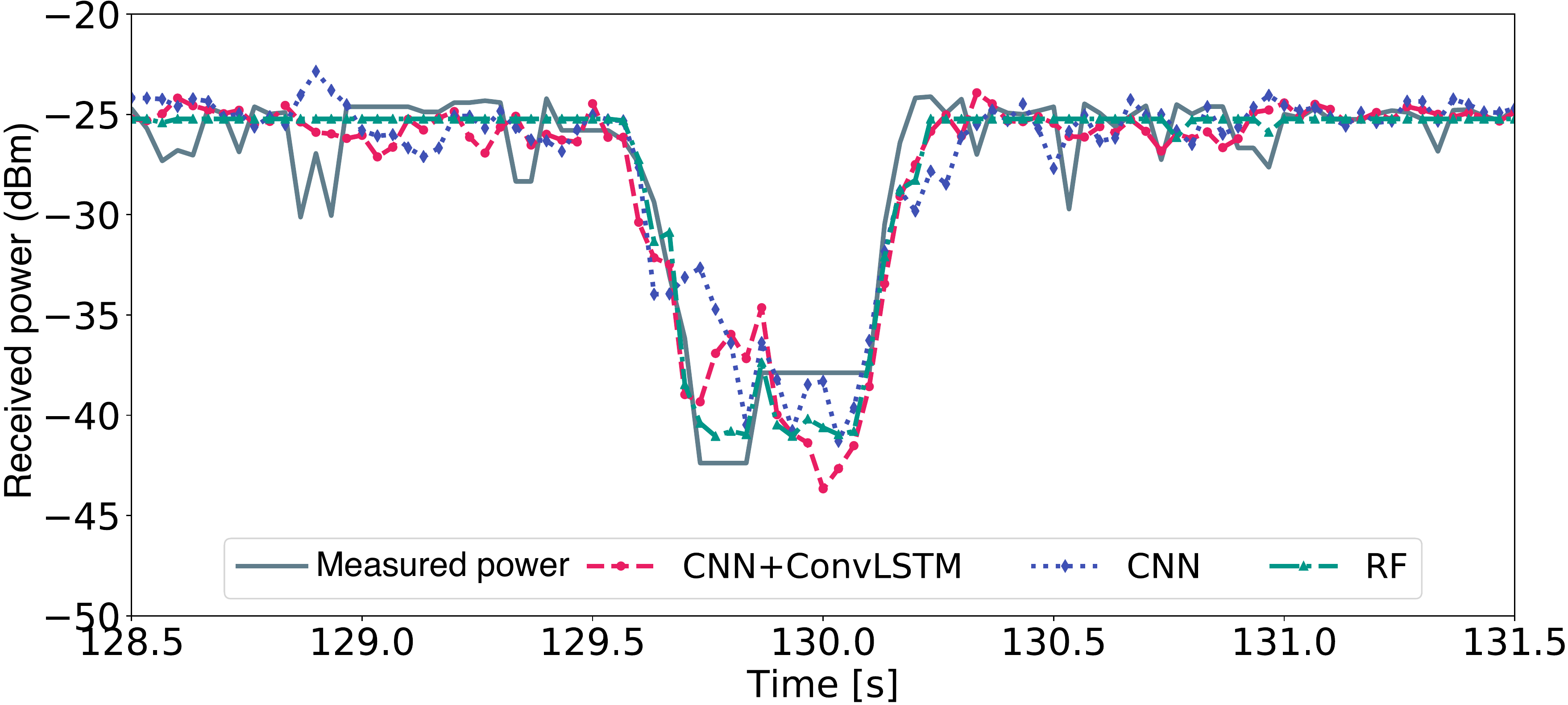}
  }\hspace{0.3em}
  \subfloat[Another portion of the predicted future received power in 500\,ms ahead when the camera was at A. In this portion, pedestrians simultaneously blocked the LOS path.\label{fig:result_time_exp_a_dbl_future}]{
      \includegraphics[width=0.45\textwidth]{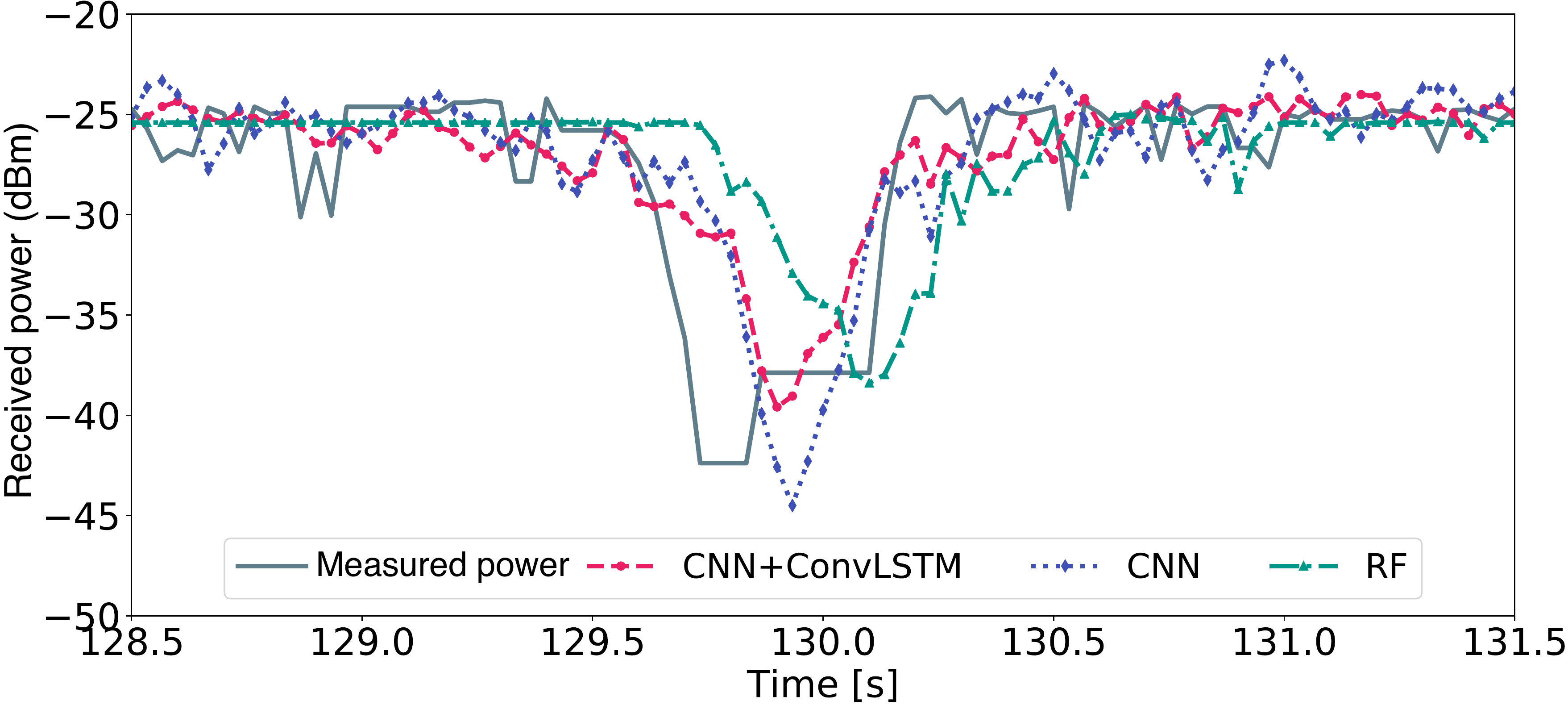}
  }
  \caption{Portions of the time-series of the measured (grand truth) and predicted received power.\label{fig:result_time_exp}}
\end{figure*}

\begin{figure}[!t]
    \centering
    \subfloat[Current received power prediction (0\,ms ahead).\label{fig:resultexp_0}]{
        \includegraphics[width=0.22\textwidth]{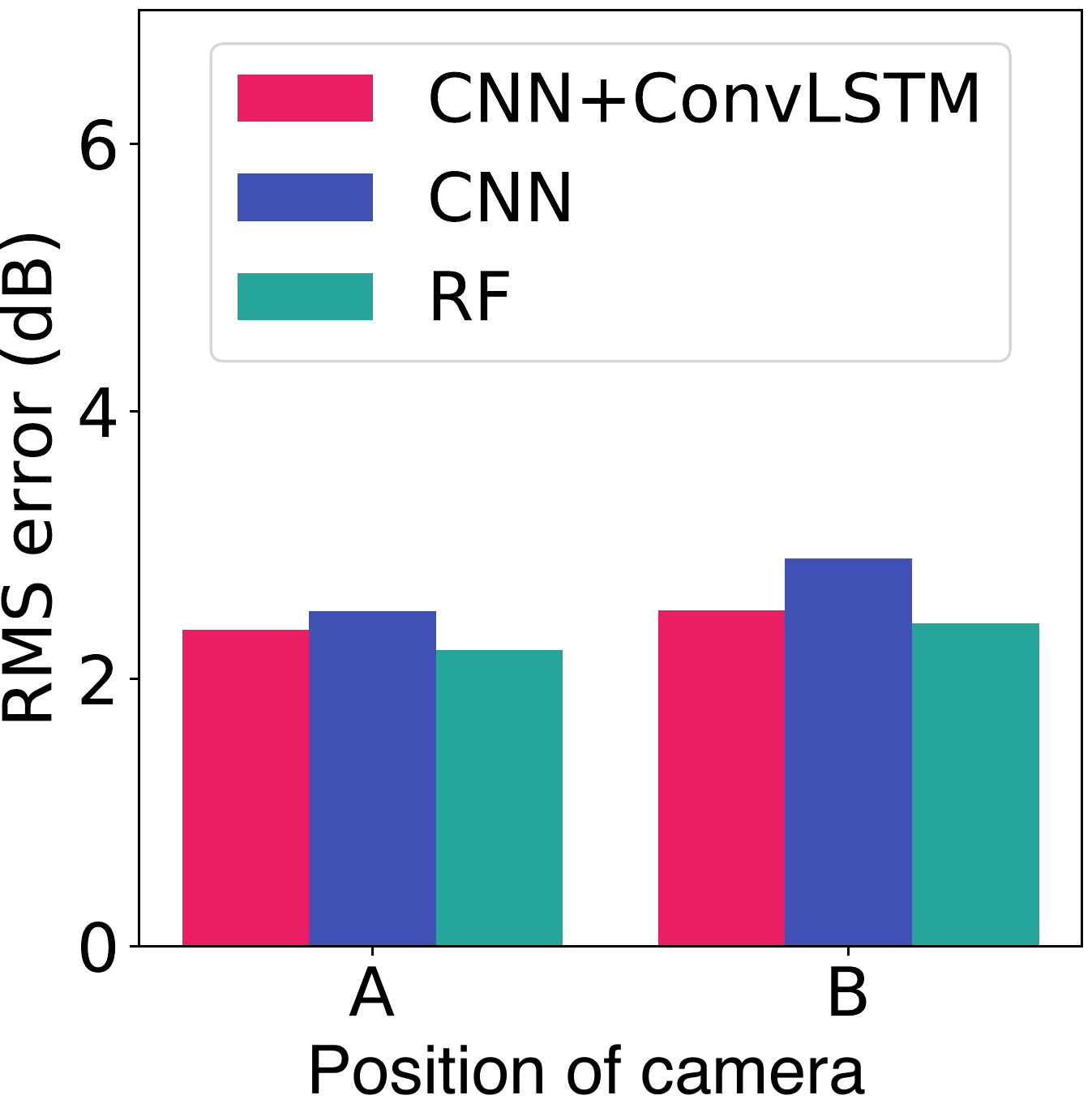}
    }\hspace{0.3em}
    \subfloat[Future received power prediction (500\,ms ahead).\label{fig:resultexp_15}]{
        \includegraphics[width=0.22\textwidth]{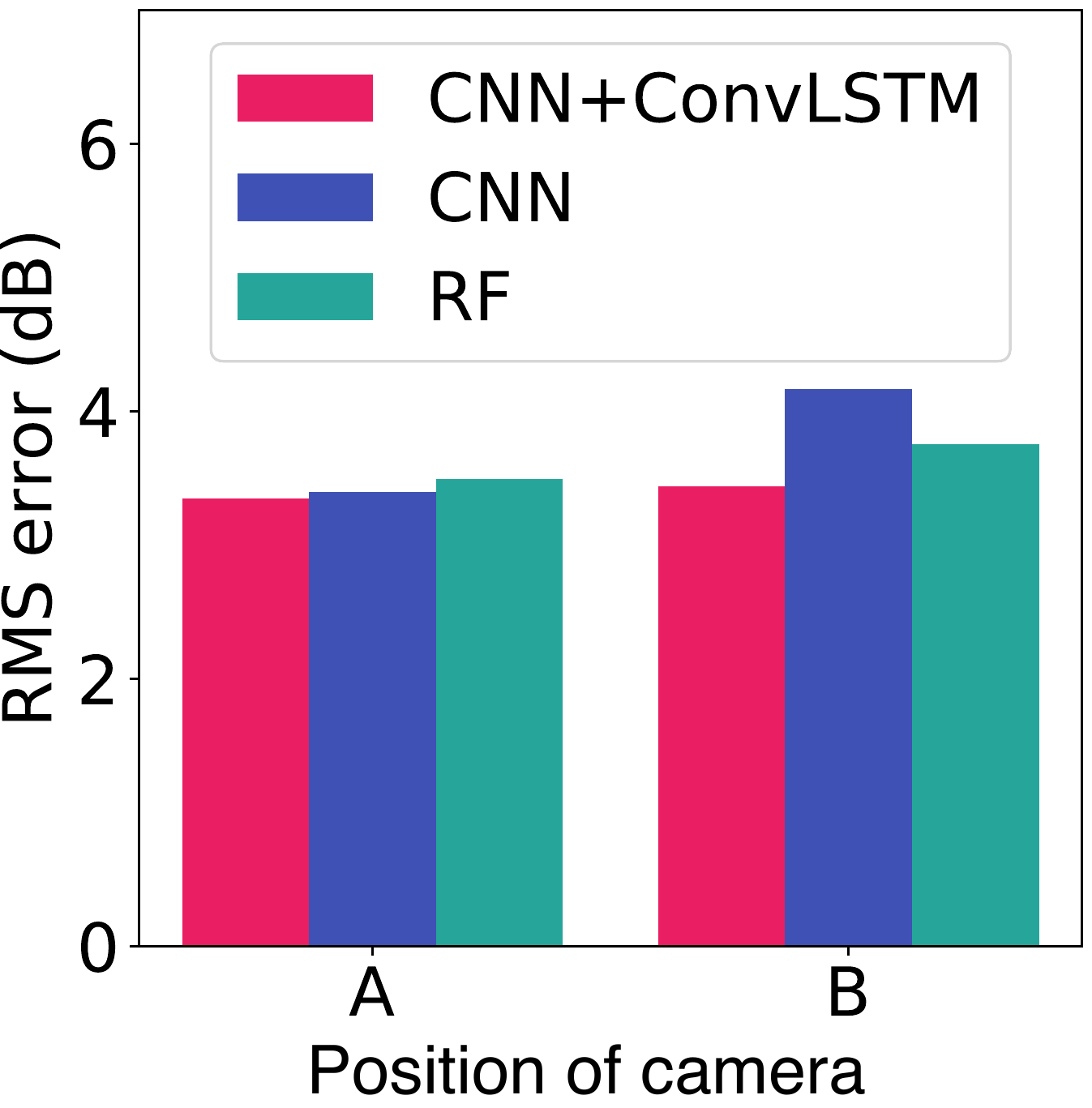}
    }
    \caption{RMS error of each algorithm. \label{fig:result_rmse_exp}}
  \end{figure}  

Fig.~\ref{fig:result_time_exp} presents the time series of the measured received power, as well as the received power predicted by each algorithm when the camera was placed at positions A or B.
The prediction results of the current and future received power matched the measured received power, in particular, CNN+ConvLSTM matched well owing to its architecture to capture spatiotemporal features.
However, in the future received power prediction, the timings when the predicted values drop were sometimes shifted by several frames from the measured values as shown in Fig.~\ref{fig:result_time_exp}\subref{fig:result_time_exp_a_dbl_future}. 
In the figure, two pedestrians walked side by side and simultaneously blocked the LOS path.
The mobility of the pedestrian behind the other was difficult to be captured since the pedestrians were overlapping in the FOV. The FOV blockage resulted in a possibly erroneous prediction of the LOS blockage timing.
It can be addressed by placing the camera at a somewhat higher position as described in Section~\ref{subsec:simulation_result}.
Moreover, the number of the data samples for the simultaneous blockage was less than that by a pedestrian. 
Generally, model training minimizes mean errors between predicted values and actual values, therefore, prediction errors for infrequent events tend to be larger than that for frequent events, which is called an imbalance problem. The learning from the imbalance data is an open issue in ML and techniques addressing the imbalance problem such oversampling and undersampling have been studied \cite{Garcia08}. We can apply the conventional techniques to our mechanism.


Fig.~\ref{fig:result_rmse_exp} presents the RMS error of each experiment. The RMS errors of all the models were less than 3.0\,dB and 4.2\,dB for the current and future predictions, respectively.
Moreover, ConvLSTM achieved even 3.5\,dB for the future prediction which was the smallest RMS error among the models.
The prediction errors for the experimental dataset were larger than that for the simulation dataset shown in Fig.~\ref{fig:result_rmse_sim}.
This is because the experimental data has a lot of uncertainties such as the wireless signal propagation, the pedestrians' mobility, and the performance of experimental devices. The uncertainties made the data samples more variational and increased the difficulty of learning an accurate prediction model.
Moreover, although the prediction results of the ConvLSTM seem to match better than that of the RF in Fig.~\ref{fig:result_time_exp}, the RMS errors for the model was almost the same as that for the RF. In the prediction results of the CNN+ConvLSTM, the predicted values in LOS communications fluctuated due to overfitting the time variation of the measured received power in LOS communications.
The fluctuation causes small errors many times and increases RMS error.
Although NNs tend to overfit due to their high flexibility, RF can avoid overfitting owing to the bagging strategy. Thus, the output of the RF tended not to fluctuate in LOS communications.

These results suggest that all the models including RF provide a sufficiently high accuracy in the predictions of the received power even for a prediction in 500\,ms.

\section{Conclusion}
\label{sec:conclusion}
This paper proposed a novel mechanism to predict a time series of a received power up to several hundred milliseconds ahead by using ML and depth images.
Specifically, the time-sequential depth images labeled with the future received power values enable the ML algorithm to learn the prediction model. Three prediction models employing CNN, CNN+ConvLSTM, and RF are constructed empirically. Their prediction accuracies were evaluated by the experiments using the simulation-based dataset and the experimental dataset.
The evaluation results showed that CNN+ConvLSTM achieved the highest accuracy among the three models and in detail it predicted the received power 500\,ms ahead with RMS errors of less than 1.0\,dB and 3.5\,dB for the simulation dataset and the experimental dataset respectively.
Furthermore, the prediction experiments confirmed that the prediction could be operated quickly and in detail the computation time is less than 3\,ms per time-sequential image.
There still exist a lot of open issues in this new challenge. Our future works include developing a method to predict the received power at a moving STA, a model transfer method for a mmWave received power prediction in other locations and situations, and a specific ML algorithm outperforming the conventional ML algorithms in the prediction problem.

\bibliographystyle{IEEEtran}
\bibliography{IEEEabrv,my_bib}

\begin{thebibliography}{10}
\providecommand{\url}[1]{#1}
\csname url@samestyle\endcsname
\providecommand{\newblock}{\relax}
\providecommand{\bibinfo}[2]{#2}
\providecommand{\BIBentrySTDinterwordspacing}{\spaceskip=0pt\relax}
\providecommand{\BIBentryALTinterwordstretchfactor}{4}
\providecommand{\BIBentryALTinterwordspacing}{\spaceskip=\fontdimen2\font plus
\BIBentryALTinterwordstretchfactor\fontdimen3\font minus
  \fontdimen4\font\relax}
\providecommand{\BIBforeignlanguage}[2]{{%
\expandafter\ifx\csname l@#1\endcsname\relax
\typeout{** WARNING: IEEEtran.bst: No hyphenation pattern has been}%
\typeout{** loaded for the language `#1'. Using the pattern for}%
\typeout{** the default language instead.}%
\else
\language=\csname l@#1\endcsname
\fi
#2}}
\providecommand{\BIBdecl}{\relax}
\BIBdecl

\bibitem{Dehos2014}
C.~Dehos, J.~Gonz{\'a}lez, A.~De~Domenico, D.~Kt{\'e}nas, and L.~Dussopt,
  ``Millimeter-wave access and backhauling: {The} solution to the exponential
  data traffic increase in 5{G} mobile communications systems?'' \emph{{IEEE}
  Commun. Mag.}, vol.~52, no.~9, pp. 88--95, Sep. 2014.

\bibitem{Sakaguchi2015}
K.~Sakaguchi \emph{et~al.}, ``Millimeter-wave evolution for 5{G} cellular
  networks,'' \emph{IEICE Trans. Commun.}, vol.~98, no.~3, pp. 388--402, Mar.
  2015.

\bibitem{Osseiran14}
A.~Osseiran \emph{et~al.}, ``{Scenarios for 5G mobile and wireless
  communications: The vision of the METIS project},'' \emph{{IEEE} Commun.
  Mag.}, vol.~52, no.~5, pp. 26--35, May 2014.

\bibitem{IMT15}
{ITU-R Recommendation}, ``{IMT} vision--framework and overall objectives of the
  future development of {IMT} for 2020 and beyond,'' \emph{Recommendation
  {M.2083}}, Sep. 2015.

\bibitem{Collonge2004}
S.~Collonge, G.~Zaharia, and G.~E. Zein, ``Influence of the human activity on
  wide-band characteristics of the 60 {GH}z indoor radio channel,''
  \emph{{IEEE} Trans. Wireless Commun.}, vol.~3, no.~6, pp. 2396--2406, Nov.
  2004.

\bibitem{yamamoto2008}
A.~Yamamoto, K.~Ogawa, T.~Horimatsu, A.~Kato, and M.~Fujise, ``Path-loss
  prediction models for intervehicle communication at 60 {GHz},'' \emph{{IEEE}
  Trans. Veh. Technol.}, vol.~57, no.~1, pp. 65--78, Jan. 2008.

\bibitem{Gao2014}
B.~Gao, Z.~Xiao, C.~Zhang, L.~Su, D.~Jin, and L.~Zeng, ``Double-link beam
  tracking against human blockage and device mobility for 60-{GH}z {WLAN},'' in
  \emph{Proc. IEEE Wireless Commun. and Networking Conf. (WCNC)}, Istanbul,
  Turkey, Apr. 2014, pp. 323--328.

\bibitem{react1}
Y.~M. Tsang and A.~S.~Y. Poon, ``Detecting human blockage and device movement
  in mm{W}ave communication system,'' in \emph{Proc. IEEE Global Commun. Conf.
  (GLOBECOM)}, Houston, Texas, USA, Dec. 2011, pp. 1--6.

\bibitem{perahia2011gigabit}
E.~Perahia and M.~X. Gong, ``Gigabit wireless {LANs}: {A}n overview of {IEEE}
  802.11ac and 802.11ad,'' \emph{ACM SIGMOBILE Mobile Comput. Commun. Rev.},
  vol.~15, no.~3, pp. 23--33, July 2011.

\bibitem{11ad_channel_model}
A.~Maltsev \emph{et~al.}, ``Channel models for 60 {GH}z {WLAN} systems, doc.:
  {IEEE} 802.11-09/0334r8,'' \emph{IEEE 802.11 09/0334r8}, July 2009.

\bibitem{nishio15}
T.~Nishio, R.~Arai, K.~Yamamoto, and M.~Morikura, ``Proactive traffic control
  based on human blockage using {RGB-D} cameras for millimeter-wave
  communications,'' in \emph{Proc. IEEE Consumer Commun. \& Networking Conf.
  (CCNC)}, Las Vegas, NV, USA, Jan. 2015, pp. 23--24.

\bibitem{oguma15}
Y.~Oguma, R.~Arai, T.~Nishio, K.~Yamamoto, and M.~Morikura, ``Proactive base
  station selection based on human blockage prediction using {RGB-D} cameras
  for {mmWave} communications,'' in \emph{Proc. IEEE Global Commun. Conf.
  (GLOBECOM)}, San Diego, CA, USA, Dec. 2015, pp. 1--6.

\bibitem{simic16}
L.~Simi{\'c}, J.~Arnold, M.~Petrova, and P.~M{\"a}h{\"a}nen, ``{RadMAC}:
  {R}adar-enabled link obstruction avoidance for agile mm-wave beamsteering,''
  in \emph{Proc. ACM 3rd Workshop on Hot Topics in Wireless (HotWireless
  2016)}, Oct. 2016, pp. 61--65.

\bibitem{ulvan13}
A.~Ulvan, R.~Bestak, and M.~Ulvan, ``Handover procedure and decision strategy
  in {LTE}-based femtocell network,'' \emph{Telecommun. Syst.}, vol.~52, no.~4,
  pp. 2733--2748, Apr. 2013.

\bibitem{satoda12}
K.~Satoda, H.~Yoshida, H.~Ito, and K.~Ozawa, ``Adaptive video pacing method
  based on the prediction of stochastic {TCP} throughput,'' in \emph{Proc. IEEE
  Global Commun. Conf. (GLOBECOM)}, Dec. 2012, pp. 1944--1950.

\bibitem{kanai16}
K.~Kanai \emph{et~al.}, ``Proactive content caching for mobile video utilizing
  transportation systems and evaluation through field experiments,''
  \emph{{IEEE} J. Sel. Areas Commun.}, vol.~34, no.~8, pp. 2102--2114, Aug.
  2016.

\bibitem{Haneda2015}
K.~Haneda, ``Channel models and beamforming at millimeter-wave frequency
  bands,'' \emph{IEICE Trans. Commun.}, vol.~98, no.~5, pp. 755--772, May 2015.

\bibitem{di2015stochastic}
M.~Di~Renzo, ``Stochastic geometry modeling and analysis of multi-tier
  millimeter wave cellular networks,'' \emph{{IEEE} Trans. Wireless Commun.},
  vol.~14, no.~9, pp. 5038--5057, Sep. 2015.

\bibitem{gapeyenko2016}
M.~Gapeyenko \emph{et~al.}, ``Analysis of human-body blockage in urban
  millimeter-wave cellular communications,'' in \emph{Proc. IEEE Int. Conf.
  Commun. (ICC)}, May 2016, pp. 1--7.

\bibitem{LQ_prediction}
L.~Liu, Y.~Fan, J.~Shu, and K.~Yu, ``A link quality prediction mechanism for
  {WSN}s based on time series model,'' in \emph{Proc. IEEE UIC/ATC}, Oct. 2010,
  pp. 175--179.

\bibitem{LQ_prediction2}
K.~Farkas, T.~Hossmann, F.~Legendre, B.~Plattner, and S.~K. Das, ``Link quality
  prediction in mesh networks,'' \emph{Comput. Commun.}, vol.~31, no.~8, pp.
  1497--1512, May 2008.

\bibitem{yao08}
J.~Yao, S.~S. Kanhere, and M.~Hassan, ``An empirical study of bandwidth
  predictability in mobile computing,'' in \emph{Proc. 3rd {ACM} Int. Workshop
  on Wireless Network Testbeds, Experimental Evaluation and Characterization},
  Sep. 2008, pp. 11--18.

\bibitem{okamoto_vtc}
H.~Okamoto, T.~Nishio, M.~Morikura, K.~Yamamoto, D.~Murayama, and K.~Nakahira,
  ``Machine-learning-based throughput estimation using images for {mmWave}
  communications,'' in \emph{Proc. IEEE Veh. Technol. Conf. (VTC2017-Spring)},
  Sydney, Australia, June 2017, pp. 1--6.

\bibitem{okamoto_ccnc}
H.~Okamoto, T.~Nishio, M.~Morikura, and K.~Yamamoto, ``Recurrent neural
  network-based received signal strength estimation using depth images for
  mmwave communications,'' in \emph{Proc. IEEE Consumer Commun. \& Networking
  Conf. (CCNC)}, Las Vegas, NV, USA, Jan. 2018, pp. 1--2.

\bibitem{zhang03}
G.~P. Zhang, ``Time series forecasting using a hybrid arima and neural network
  model,'' \emph{Neurocomputing}, vol.~50, pp. 159--175, Jan. 2003.

\bibitem{RNN_SNR}
J.~Zhang and I.~Marsic, ``Link quality and signal-to-noise ratio in 802.11
  {WLAN} with fading: {A} time-series analysis,'' in \emph{Proc. IEEE Veh.
  Technol. Conf. (VTC2006-Fall)}, Montreal, Quebec, Canada, Sep. 2006, pp.
  1--5.

\bibitem{long08}
X.~Long and B.~Sikdar, ``A real-time algorithm for long range signal strength
  prediction in wireless networks,'' in \emph{Proc. IEEE Wireless Commun. and
  Networking Conf. (WCNC)}, Apr. 2008, pp. 1120--1125.

\bibitem{ray_tracing_11ad}
M.~Jacob, S.~Priebe, A.~Maltsev, A.~Lomayev, V.~Erceg, and T.~K{\"u}rner, ``A
  ray tracing based stochastic human blockage model for the {IEEE} 802.11ad 60
  {GHz} channel model,'' in \emph{Proc. 5th European Conf. Antennas and
  Propagation (EuCAP)}, Rome, Italy, Apr. 2011, pp. 3084--3088.

\bibitem{ray_tracing_mmwave_calc}
V.~Degli-Esposti \emph{et~al.}, ``Ray-tracing-based mm-wave beamforming
  assessment,'' \emph{IEEE Access}, vol.~2, pp. 1314--1325, Oct. 2014.

\bibitem{double_knife_edge}
G.~R. MacCartney, S.~Deng, S.~Sun, and T.~S. Rappaport, ``Millimeter-wave human
  blockage at 73 {GH}z with a simple double knife-edge diffraction model and
  extension for directional antennas,'' in \emph{Proc. IEEE Veh. Technol. Conf.
  (VTC2016-Fall)}, Montreal, Canada, Sep. 2016, pp. 1--6.

\bibitem{11ad}
``{Wireless LAN Medium Access Control (MAC) and Physical Layer (PHY)
  Specifications Amendment 3: Enhancements for Very High Throughput in the 60
  {GHz} Band},'' IEEE Std. 802.11ad, IEEE 802.11 Working Group, 2012.

\bibitem{oguma2016proactive}
Y.~Oguma, T.~Nishio, K.~Yamamoto, and M.~Morikura, ``Proactive handover based
  on human blockage prediction using {RGB-D} cameras for mm{W}ave
  communications,'' \emph{IEICE Trans. Commun.}, vol.~99, no.~8, pp.
  1734--1744, Oct. 2016.

\bibitem{conv3d}
S.~Ji, W.~Xu, M.~Yang, and K.~Yu, ``{3D} convolutional neural networks for
  human action recognition,'' \emph{{IEEE} Trans. Pattern Anal. Mach. Intell.},
  vol.~35, no.~1, pp. 221--231, Jan. 2013.

\bibitem{biswas2012depth}
J.~Biswas and M.~Veloso, ``Depth camera based indoor mobile robot localization
  and navigation,'' in \emph{Proc. IEEE Int. Conf. Robotics and Automation
  (ICRA)}, St Paul, MN, USA, May 2012, pp. 1697--1702.

\bibitem{kinect}
Kinect, https://dev.windows.com{\slash}en-us{\slash}kinect, Nov. 2017.

\bibitem{lazaros2008review}
N.~Lazaros, G.~C. Sirakoulis, and A.~Gasteratos, ``Review of stereo vision
  algorithms: From software to hardware,'' \emph{Int. J. Optomechatronics},
  vol.~2, no.~4, pp. 435--462, Nov. 2008.

\bibitem{convLSTM}
X.~Shi, Z.~Chen, H.~Wang, D.~Yeung, W.~Wong, and W.~Woo, ``Convolutional {LSTM}
  network: {A} machine learning approach for precipitation nowcasting,'' in
  \emph{Proc. Neural Inform. Process. Syst. (NIPS)}, Montreal, Canada, Dec.
  2015, pp. 802--810.

\bibitem{mnih2015}
V.~Mnih \emph{et~al.}, ``Human-level control through deep reinforcement
  learning,'' \emph{Nature}, vol. 518, no. 7540, p. 529, Feb. 2015.

\bibitem{cnn_survey}
J.~Gu \emph{et~al.}, ``Recent advances in convolutional neural networks,''
  \emph{Pattern Recogn.}, vol.~77, no.~C, pp. 354--377, May 2018.

\bibitem{LSTM}
S.~Hochreiter and J.~Schmidhuber, ``Long {S}hort-{T}erm {M}emory,''
  \emph{Neural Comput.}, vol.~9, no.~8, pp. 1735--1780, Nov. 1997.

\bibitem{video_classification}
J.~Donahue, L.~A.~Hendricks, S.~Guadarrama, M.~Rohrbach, S.~Venugopalan,
  K.~Saenko, and T.~Darrell, ``Long-term recurrent convolutional networks for
  visual recognition and description,'' in \emph{Proc. Conf. Comput. Vision and
  Pattern Recognition (CVPR)}, Boston, MA, USA, June 2015, pp. 2625--2634.

\bibitem{random_forest}
L.~Breiman, ``Random forests,'' \emph{Mach. Learning}, vol.~45, no.~1, pp.
  5--32, Oct. 2001.

\bibitem{bagging}
------, ``Bagging predictors,'' \emph{Mach. Learning}, vol.~24, no.~2, pp.
  123--140, Aug. 1996.

\bibitem{Shotton2011}
J.~Shotton, A.~Fitzgibbon, M.~Cook, T.~Sharp, M.~Finocchio, R.~Moore,
  A.~Kipman, and A.~Blake, ``Real-time human pose recognition in parts from
  single depth images,'' in \emph{Proc. Conf. Comput. Vision and Pattern
  Recognition (CVPR)}, Colorado Springs, USA, June 2011, pp. 1297--1304.

\bibitem{batch_normalization}
S.~Ioffe and C.~Szegedy, ``Batch normalization: {A}ccelerating deep network
  training by reducing internal covariate shift,'' in \emph{Proc. Int. Conf.
  Machine Learning (ICML)}, vol.~37, Lille, France, July 2015, pp. 448--456.

\bibitem{keras}
Keras, https://keras.io/, Nov. 2017.

\bibitem{tensorflow}
TensorFlow, https://www.tensorflow.org/, Nov. 2017.

\bibitem{sklearn}
F.~Pedregosa \emph{et~al.}, ``Scikit-learn: Machine learning in python,''
  \emph{J. Mach. Learning Res.}, vol.~12, pp. 2825--2830, Oct. 2011.

\bibitem{nadam}
I.~Sutskever, J.~Martens, G.~Dahl, and G.~Hinton, ``On the importance of
  initialization and momentum in deep learning,'' in \emph{Proc. Int. Conf.
  Machine Learning (ICML)}, Atlanta, GA, USA, June 2013, pp. 1139--1147.

\bibitem{twin_cylinder}
T.~Wang, M.~Umehira, H.~Otsu, S.~Takeda, T.~Miyajima, and K.~Kagoshima, ``A
  twin cylinder model for moving human body shadowing in 60{GHz} {WLAN},'' in
  \emph{Proc. 21st Asia-Pacific Conf. Commun. (APCC 2015)}, Kyoto, Japan, Oct.
  2015, pp. 188--192.

\bibitem{pedestrian_poisson}
V.~Vukadinovi\'{c}, \'{O}lafur Ragnar~Helgason, and G.~Karlsson, ``An
  analytical model for pedestrian content distribution in a grid of streets,''
  \emph{Math. Comput. Model.}, vol.~57, no.~11, pp. 2933--2944, June 2013.

\bibitem{human_phantoms}
C.~Gustafson and F.~Tufvesson, ``Characterization of 60 {GHz} shadowing by
  human bodies and simple phantoms,'' in \emph{Proc. 6th European Conf.
  Antennas and Propagation (EuCAP)}, Prague, Czech Republic, Mar. 2012, pp.
  473--477.

\bibitem{RapLab}
{Kozo Keikaku Engineering, Inc.}, ``Raplab,''
  http://www4.kke.co.jp{\slash}raplab/, Nov. 2017.

\bibitem{blender}
Blender, https://www.blender.org/, Nov. 2017.

\bibitem{koda18}
Y.~Koda, K.~Yamamoto, T.~Nishio, and M.~Morikura, ``Measurement method of
  temporal attenuation by human body in off-the-shelf 60 {GHz} {WLAN} with
  {HMM}-based transmission state estimation,'' \emph{Wireless Commun. Mobile
  Comput.}, vol. 2018, Apr. 2018.

\bibitem{Garcia08}
E.~A. Garcia and H.~He, ``Learning from imbalanced data,'' \emph{{IEEE} Trans.
  Knowl. Data Eng.}, vol.~21, pp. 1263--1284, Dec. 2008.

\end{thebibliography}


\vspace{-10mm}
\begin{IEEEbiography}
	[{\includegraphics[width=1in,height=1.25in,clip,keepaspectratio]{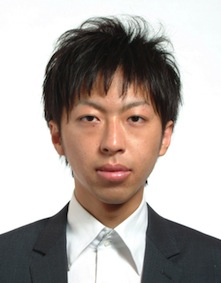}}]
	{Takayuki Nishio}
	received the B.E.\ degree in Electrical and Electronic Engineering from Kyoto University in 2010.
	He received the master and Ph.D.\ degrees in Informatics from Kyoto University in 2012 and 2013, respectively.
	He is currently an Assistant Professor in Communications and Computer Engineering, Graduate School of Informatics, Kyoto University.
	His current research interests include machine learning-based network control, machine learning in wireless networks, and heterogeneous resource management.
	He is a member of the IEEE.
\end{IEEEbiography}
\vspace{-10mm}
\begin{IEEEbiography}
	[{\includegraphics[width=1in,height=1.25in,clip,keepaspectratio]{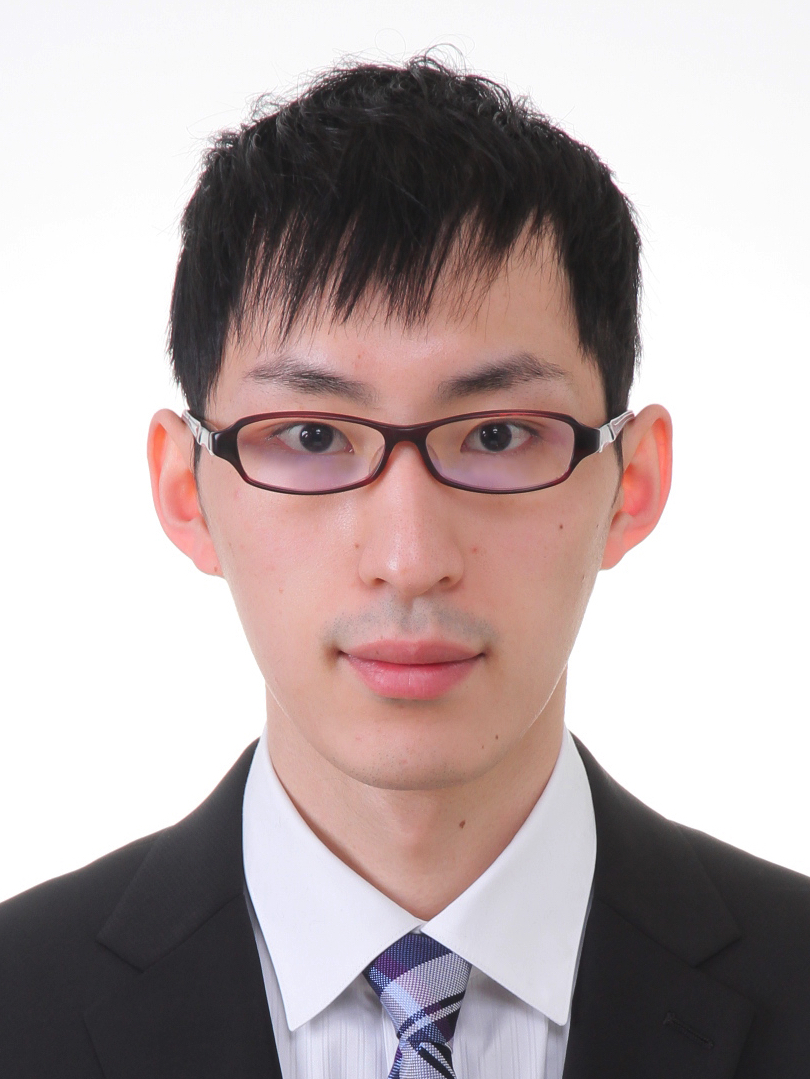}}]
	{Hironao Okamoto}
    received the B.E.\ degree in electrical and electronic engineering from Kyoto University in 2016.
    He received the master degree in Communications and Computer Engineering, Graduate School of Informatics from Kyoto University in 2018.
\end{IEEEbiography}
\vspace{-10mm}
\begin{IEEEbiography}
	[{\includegraphics[width=1in,height=1.25in,clip,keepaspectratio]{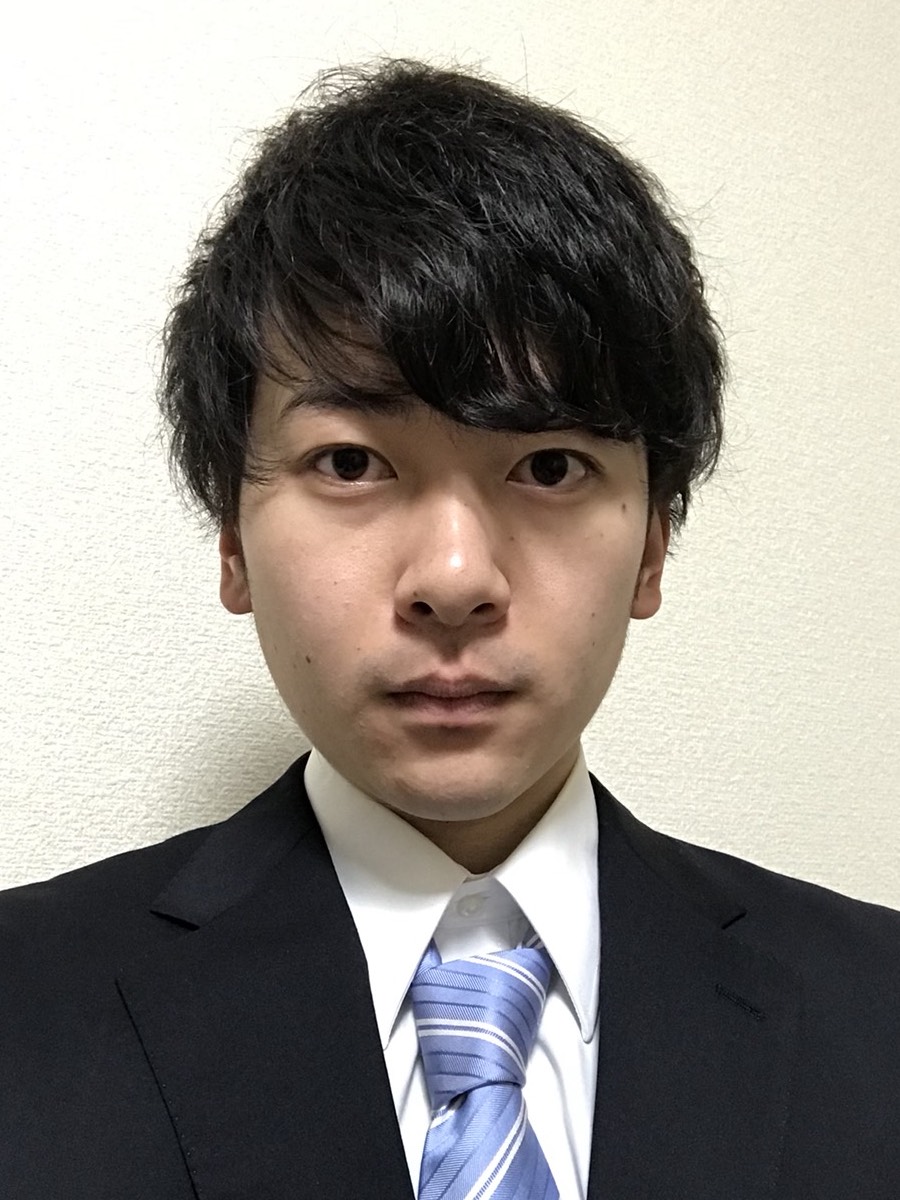}}]
	{Kota Nakashima}
	received the B.E.\@ degree in electrical and electronic engineering from Kyoto University in 2018.
	He is currently studying toward the M.E.\ degree at the Graduate School of Informatics from Kyoto University.
    He is a member of the IEICE and a member of the IEEE.
\end{IEEEbiography}
\vspace{-10mm}
\begin{IEEEbiography}
	[{\includegraphics[width=1in,height=1.25in,clip,keepaspectratio]{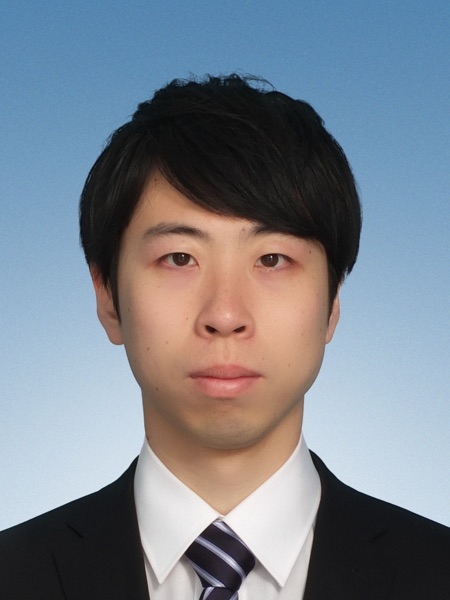}}]
	{Yusuke Koda}
	received the B.E.\@ degree in electrical and electronic engineering from Kyoto University in 2016
	and the M.E.\@ degree at the Graduate School of Informatics from Kyoto Univiesity.
	He is currently studying toward the Ph.D.\@ degree at the Graduate School of Informatics from Kyoto University.
	He received the VTS Japan Young Researcher's Encouragement Award in 2017.
	He is a member of the IEICE and a member of the IEEE.
\end{IEEEbiography}
\vspace{-10mm}
\begin{IEEEbiography}
	[{\includegraphics[width=1in,height=1.25in,clip,keepaspectratio]{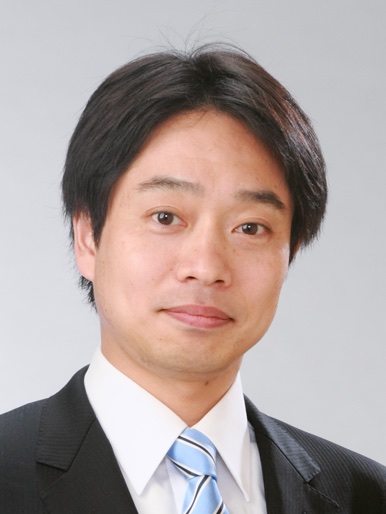}}]
	{Koji Yamamoto}
	(S'03--M'06) received the B.E. degree in electrical and electronic engineering from Kyoto University in 2002, and the M.E. and Ph.D. degrees in Informatics from Kyoto University in 2004 and 2005, respectively.
	From 2004 to 2005, he was a research fellow of the Japan Society for the Promotion of Science (JSPS).
	Since 2005, he has been with the Graduate School of Informatics, Kyoto University, where he is currently an associate professor.
	From 2008 to 2009, he was a visiting researcher at Wireless@KTH, Royal Institute of Technology (KTH) in Sweden.
	He serves as an editor of IEEE Wireless Communications Letters from 2017 and the Track Co-Chairs of APCC 2017 and CCNC 2018.
	His research interests include radio resource management and applications of game theory.
	He received the PIMRC 2004 Best Student Paper Award in 2004, the Ericsson Young Scientist Award in 2006.
	He also received the Young Researcher's Award, the Paper Award, SUEMATSU-Yasuharu Award from the IEICE of Japan in 2008, 2011, and 2016, respectively, and IEEE Kansai Section GOLD Award in 2012.
\end{IEEEbiography}
\vspace{-10mm}
\begin{IEEEbiography}
	[{\includegraphics[width=1in,height=1.25in,clip,keepaspectratio]{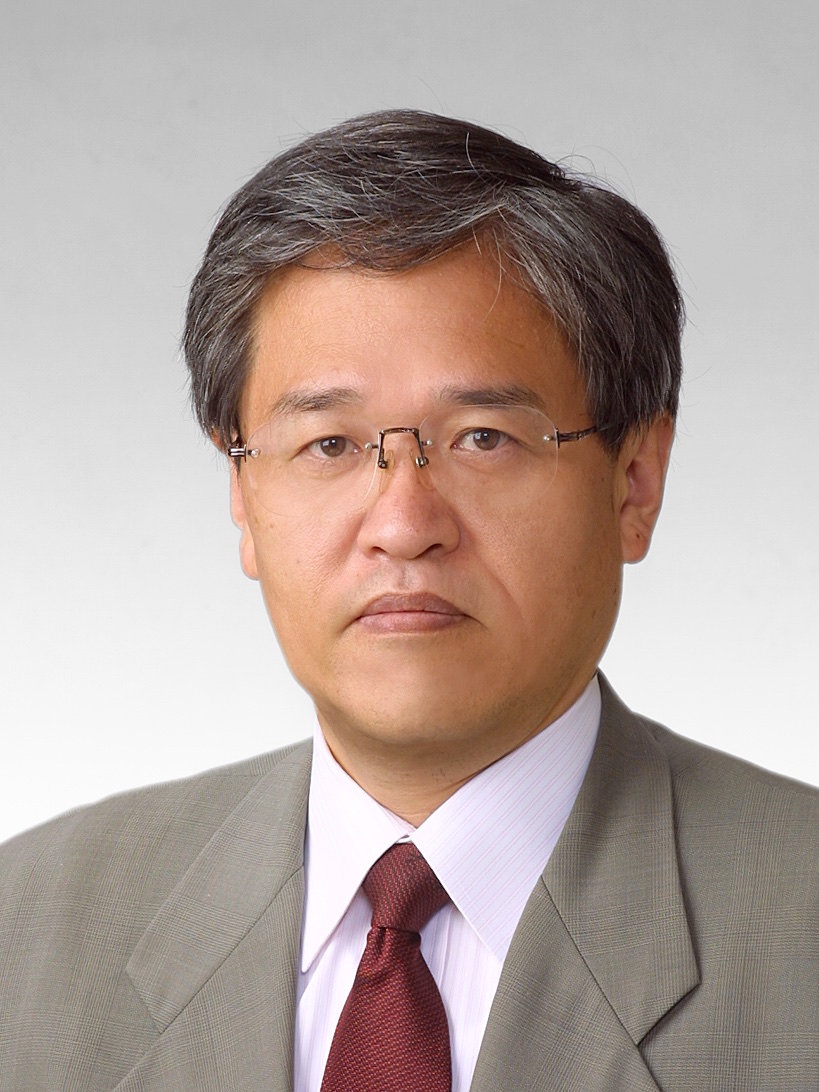}}]
	{Masahiro Morikura}
	received his B.E., M.E., and Ph.D.\ degrees in electronics engineering from Kyoto University, Kyoto, Japan in 1979, 1981 and 1991, respectively.
	He joined NTT in 1981, where he was engaged in the research and development of TDMA equipment for satellite communications. From 1988 to 1989,
	he was with the Communications Research Centre, Canada, as a guest scientist.
	From 1997 to 2002, he was active in the standardization of the IEEE 802.11a based wireless LAN.
	His current research interests include WLANs and M2M wireless systems.
	He received the Paper Award and the Achievement Award from IEICE in 2000 and 2006, respectively.
	He also received the Education, Culture, Sports, Science and Technology Minister Award in 2007 and Maejima Award in 2008,
	and the Medal of Honor with Purple Ribbon from Japan's Cabinet Office in 2015.
	Dr. Morikura is now a professor in the Graduate School of Informatics, Kyoto University. He is a member of the IEEE.
\end{IEEEbiography}
\vspace{-10mm}
\begin{IEEEbiography}
	[{\includegraphics[width=1in,height=1.25in,clip,keepaspectratio]{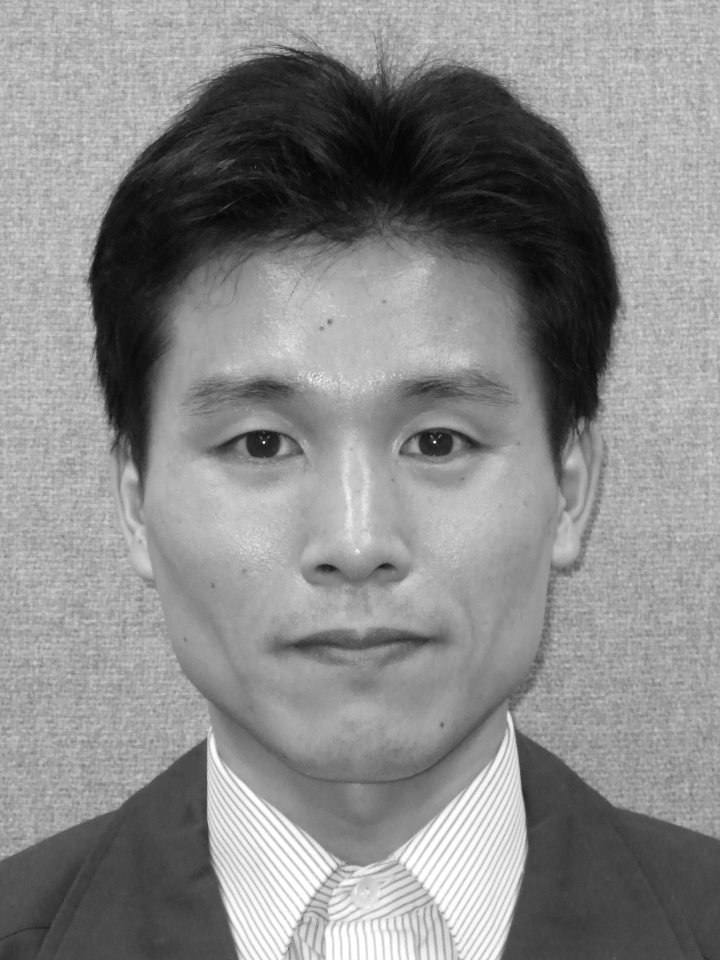}}]
	{Yusuke Asai}
	received the B.E., M.E.\ and Ph.D.\ degrees in Nagoya University, Japan, in 1997, 1999 and 2017, respectively. In 1999, he joined Nippon Telegraph and Telephone Corporation (NTT), in Japan. He has been engaged in the research and development of signal processing and resource management techniques for broadband wireless LAN systems. He has served as one of the co-chairpersons of the coexistence ad-hoc group of Task Group ac in IEEE 802.11 Working Group. He is currently a senior research engineer in NTT Network Innovation Laboratories. He received the Young Researcher's Award of the IEICE Japan in 2004 and the Certification of Appreciation for outstanding contributions to the development of IEEE 802.11ac-2013 from the IEEE-SA in 2014. He is a member of IEEE.
\end{IEEEbiography}
\vspace{-10mm}
\begin{IEEEbiography}
	[{\includegraphics[width=1in,height=1.25in,clip,keepaspectratio]{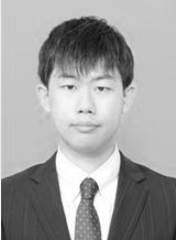}}]
	{Ryo Miyatake}
	received the B.E., M.E.\ degrees in University of Electro-Communications, Japan, in 2011 and 2013, respectively. In 2013, he joined Nippon Telegraph and Telephone Corporation (NTT), in Japan. He has been engaged in the research and development of signal processing techniques for wireless communication systems. He is currently a research engineer in NTT Network Innovation Laboratories.
\end{IEEEbiography}
\end{document}